\documentclass[12pt,english]{article}

\usepackage{amsmath}
\usepackage{amssymb}
\usepackage{graphicx}
\usepackage{cite}

\makeatletter

 \@addtoreset{equation}{section}

\allowdisplaybreaks[1]

\newif\ifContLineOne
\newif\ifContLineTwo
\newif\ifContLineThree

\def\conC#1{\vbox{\ialign{##\crcr
  \ifContLineThree\hrulefill\else\vphantom{\hrulefill}\fi\crcr
  \noalign{\kern3.2pt\nointerlineskip}
  \ifContLineTwo\hrulefill\else\vphantom{\hrulefill}\fi\crcr
  \noalign{\kern3.2pt\nointerlineskip}
  \ifContLineOne\hrulefill\else\vphantom{\hrulefill}\fi\crcr
  \noalign{\nointerlineskip}
  $\hfil\textstyle{\vbox to 14pt{}#1}\hfil$\crcr}}}

\def\DrawLeg#1#2{
  \kern-.2pt              
  \dimen2 =#1             
  \advance\dimen2 by 2pt  
  \dimen3 = 10.6pt        
  \dimen4 =3.6pt          
  \advance\dimen3 by -\dimen2 
  \multiply\dimen4 by #2
  \advance\dimen3 by \dimen4
  \raise\dimen2 \hbox{\vrule height\dimen3 width .4pt} 
  \kern-.2pt}             

\def\begC#1#2{\setbox0 =\hbox{$\textstyle{#2}$}
  \dimen0=.5\wd0 \dimen1=\ht0
  \conC{\hskip\dimen0}
  \count255=#1
  \ifnum\count255 =1 \ContLineOnetrue\else
  \ifnum\count255 =2 \ContLineTwotrue\else
  \ifnum\count255 =3 \ContLineThreetrue\fi\fi\fi
  \DrawLeg{\dimen1}{\count255}
  \conC{\hskip\dimen0}
  \kern-\dimen0\kern-\dimen0 \box0}

\def\endC#1#2{\setbox0 =\hbox{$\textstyle{#2}$}
  \dimen0=.5\wd0 \dimen1=\ht0
  \conC{\hskip\dimen0}
  \count255=#1
  \ifnum\count255 =1 \ContLineOnefalse\else
  \ifnum\count255 =2 \ContLineTwofalse\else
  \ifnum\count255 =3 \ContLineThreefalse\fi\fi\fi
  \DrawLeg{\dimen1}{\count255}
  \conC{\hskip\dimen0}
  \kern-\dimen0\kern-\dimen0 \box0}

\makeatother

\usepackage{babel}

\begin{document}
\begin{titlepage}

\global\long\def\thefootnote{\fnsymbol{footnote}}

\begin{flushright}
\begin{tabular}{l}
UTHEP-693 \tabularnewline
\end{tabular}
\par\end{flushright}

\bigskip{}

\begin{center}
\textbf{\Large{}Multiloop Amplitudes of Light-cone Gauge NSR String
Field Theory in Noncritical Dimensions}{\Large{} }
\par\end{center}{\Large \par}

\bigskip{}

\begin{center}
{\large{}{}Nobuyuki Ishibashi}$^{a}$\footnote{e-mail: ishibash@het.ph.tsukuba.ac.jp}
{\large{}{}and Koichi Murakami}$^{b}$\footnote{e-mail: koichi@kushiro-ct.ac.jp} 
\par\end{center}

\begin{center}
$^{a}$\textit{Center for Integrated Research in Fundamental Science
and Engineering (CiRfSE),}\\
\textit{ Faculty of Pure and Applied Sciences, University of Tsukuba}\\
\textit{ Tsukuba, Ibaraki 305-8571, JAPAN}\\
 
\par\end{center}

\begin{center}
$^{b}$\textit{National Institute of Technology, Kushiro College,}\\
\textit{ Otanoshike-Nishi 2-32-1, Kushiro, Hokkaido 084-0916, JAPAN} 
\par\end{center}

\bigskip{}

\bigskip{}

\bigskip{}

\begin{abstract}
Feynman amplitudes of light-cone gauge superstring field theory are
ill-defined because of various divergences. In a previous paper, one
of the authors showed that taking the worldsheet theory to be the
one in a linear dilaton background $\Phi=-iQX^{1}$ with Feynman $i\varepsilon$
$(\varepsilon>0)$ and $Q^{2}>10$ yields finite amplitudes. In this
paper, we apply this worldsheet theory to dimensional regularization
of the light-cone gauge NSR superstring field theory. We concentrate
on the amplitudes for even spin structure with external lines in the
(NS,NS) sector. We show that the multiloop amplitudes are indeed regularized
in our scheme and that they coincide with the results in the first-quantized
formalism through the analytic continuation $Q\to0$. 
\end{abstract}
\global\long\def\thefootnote{\arabic{footnote}}

\end{titlepage}

\pagebreak{}

\section{Introduction}

Since the light-cone gauge superstring field theory \cite{Mandelstam:1974hk,Mandelstam:1985wh,Sin:1988yf,Green1983b,Green1983,Gross1987}
takes a simple form, this theory is expected to be very useful in
studying superstring theory. In a series of papers \cite{Baba:2009kr,Baba:2009ns,Baba:2009fi,Baba:2009zm,Ishibashi:2010nq,Ishibashi:2011fy,Ishibashi:2013nma,Ishibashi2016b},
using the light-cone gauge closed NSR superstring field theory, we
have studied the contact term divergences \cite{Greensite:1986gv,Greensite:1987sm,Greensite:1987hm,Green:1987qu,Wendt:1987zh}.
We have pointed out that the contact term divergences can be regularized
by shifting the central charge of the light-cone gauge worldsheet
theory to a sufficiently large negative value \cite{Baba:2009kr}.
We refer to this type of regularization as the dimensional regularization,
since the central charge is directly related to the spacetime dimensions
in string theory. We have considered mainly two ways to shift the
central charge in the regularization: The one is to naively shift
the number of the transverse dimensions $d-2$; The other is to add
a superconformal field theory with central charge large negative to
the worldsheet theory.

Recently, one of the authors has proposed another prescription \cite{Ishibashi2016a},
in which the string theory in a linear dilaton background $\Phi=-iQX^{1}$
is considered so that the central charge of the system becomes $12-12Q^{2}$.
In \cite{Ishibashi2016a}, the divergences which appear in the amplitudes
of this theory have been thoroughly studied. It has been shown that
the Feynman amplitudes of light-cone gauge superstring field theory
in the linear dilaton background are indeed finite, when the theory
is with the Feynman $i\varepsilon$ $(\varepsilon>0)$ and $Q^{2}>10$.

In this paper, we use this theory to dimensionally regularize and
calculate the Feynman amplitudes. In this work, we restrict ourselves
to the amplitudes for even spin structure with external lines in the
(NS,NS) sector for simplicity. In this scheme, we define the amplitudes
as analytic functions of $Q$ and perform the analytic continuation
$Q\to0$. We show that the limit $Q\to0$ is smooth except the divergences
coming from the boundaries of the moduli space and the results coincide
with those from the first-quantized method.\footnote{In \cite{Sen2016,Sen2016c,Sen2016d}, Sen has constructed covariant
string field theories which reproduce the Feynman amplitudes from
the first-quantized approach. } In order to show the coincidence between our results and those in
the first-quantization, we recast the amplitudes into a BRST invariant
form in terms of the conformal gauge worldsheet theory. For this purpose,
together with the superreparametrization ghosts, we introduce the
longitudinal variables with nonstandard interactions, which is the
supersymmetric $X^{\pm}$ CFT constructed in \cite{Baba:2009fi} with
the identification ${\displaystyle \frac{d-10}{8}}=-Q^{2}$.

The organization of this paper is as follows. In section \ref{sec:Superstring-field-theory},
we recapitulate the light-cone gauge superstring field theory in the
linear dilaton background constructed in \cite{Ishibashi2016a}. In
section \ref{sec:BRST-invariant-form}, we show that the multiloop
amplitudes can be expressed in terms of a BRST invariant worldsheet
theory in the conformal gauge. For this purpose, we add the supersymmetric
$X^{\pm}$ CFT and superreparametrization ghosts to the worldsheet
theory. We show that the supercurrents in the light-cone gauge, inserted
at the interaction points, can be transformed into the picture changing
operators (PCO's) in the conformal gauge and the expressions become
BRST invariant. In section \ref{sec:The-amplitudes-from}, we carry
out the analytic continuation $Q\to0$ of the Feynman amplitudes.
We show that the results from the first-quantized formalism are reproduced.
Namely, the results obtained here coincide with those obtained using
the Sen-Witten prescription \cite{Sen2015a,Sen2015b,Sen2015}, up
to infrared divergence problems. Section \ref{sec:Conclusions-and-discussions}
is devoted to conclusions and discussions. In appendix \ref{sec:Arakelov-metric-and},
the definitions of the Arakelov metric and Arakelov Green's functions
are presented. In appendix \ref{sec:Supersymmetric--CFT}, some details
of the supersymmetric $X^{\pm}$ CFT are given. Formulas used in the
text are provided in appendices \ref{sec:-systems-on-higher} and
\ref{sec:A-proof-of}.

\section{Superstring field theory in linear dilaton background\label{sec:Superstring-field-theory}}

In this section, we review the light-cone gauge superstring field
theory in linear dilaton background constructed in \cite{Ishibashi2016a}.
The string field theory is given for Type II superstring theory formulated
in the NSR formalism. The heterotic case can be dealt with in a similar
way.

\subsection{Linear dilaton background\label{sec:Worldsheet-theory}}

In order to regularize various divergences, we consider the superstring
theory in a linear dilaton background $\Phi=-iQX^{1}$, with a real
constant $Q$. The worldsheet action of $X^{1}$ and its fermionic
partners $\psi^{1},\bar{\psi^{1}}$ on a worldsheet with metric $ds^{2}=2\hat{g}_{z\bar{z}}dzd\bar{z}$
becomes 
\begin{eqnarray}
S\left[X^{1},\psi^{1},\bar{\psi}^{1};\hat{g}_{z\bar{z}}\right] & = & \frac{1}{8\pi}\int dz\wedge d\bar{z}\sqrt{\hat{g}}\left(\hat{g}^{ab}\partial_{a}X^{1}\partial_{b}X^{1}-2iQ\hat{R}X^{1}\right)\nonumber \\
 &  & \qquad+\frac{1}{4\pi}\int dz\wedge d\bar{z}i\left(\psi^{1}\bar{\partial}\psi^{1}+\bar{\psi}^{1}\partial\bar{\psi}^{1}\right)\,,\label{eq:linaction}
\end{eqnarray}
and the energy-momentum tensor and the supercurrent, which generate
the superconformal transformations, are given as 
\begin{eqnarray}
T^{X^{1}}(z) & = & -\frac{1}{2}(\partial X^{1})^{2}-iQ(\partial-\partial\ln\hat{g}_{z\bar{z}})\partial X^{1}-\frac{1}{2}\psi^{1}\partial\psi^{1}\nonumber \\
 &  & \ -Q^{2}\left(-\frac{1}{2}\left(\partial\ln g_{z\bar{z}}\right)^{2}+\partial^{2}\ln g_{z\bar{z}}\right),\nonumber \\
T_{F}^{X^{1}}(z) & = & -\frac{i}{2}\partial X^{1}\psi^{1}+Q(\partial-\frac{1}{2}\partial\ln\hat{g}_{z\bar{z}})\psi^{1}\,.
\end{eqnarray}

In order to construct string field theory and calculate amplitudes
we need the correlation functions of the linear dilaton conformal
field theory. Since the fermionic part is just a free theory, we concentrate
on the bosonic part. Defining 
\begin{equation}
\tilde{X}^{1}\equiv X^{1}-iQ\ln(2g_{z\bar{z}})\,,
\end{equation}
the correlation function of $e^{ip_{r}\tilde{X}^{1}}(Z_{r},\bar{Z}_{r})\,(r=1,\cdots,N)$
can be calculated on a Riemann surface \cite{Ishibashi2016a}. Using
the Arakelov metric $ds^{2}=2g_{z\bar{z}}^{\mathrm{A}}dzd\bar{z}$
\cite{arakelov} defined on the surface, it is given as 
\begin{eqnarray}
\lefteqn{\int\left[dX^{1}\right]_{g_{z\bar{z}}}e^{-S\left[X^{1};g_{z\bar{z}}\right]}\prod_{r=1}^{N}e^{ip_{r}\tilde{X}^{1}}(Z_{r},\bar{Z}_{r})}\nonumber \\
 &  & =2\pi\delta\left(\sum_{r}p_{r}+2Q(1-g)\right)e^{-\frac{1-12Q^{2}}{24}\Gamma\left[\sigma;g_{z\bar{z}}^{\mathrm{A}}\right]}Z^{X}\left[g_{z\bar{z}}^{\mathrm{A}}\right]\prod_{r>s}e^{-p_{r}p_{s}G^{\mathrm{A}}(Z_{r},Z_{s})}\prod_{r}\left(2g_{z\bar{z}}^{\mathrm{A}}\right)^{\frac{1}{2}p_{r}^{2}+Qp_{r}},\nonumber \\
 &  & \ \label{eq:linearcorr}
\end{eqnarray}
where $Z^{X}\left[g_{z\bar{z}}^{\mathrm{A}}\right]$ denotes the partition
function for a free boson on the worldsheet with the Arakelov metric
(\ref{eq:ZXg}), and 
\begin{eqnarray}
 &  & S\left[X^{1};g_{ab}\right]=\frac{1}{8\pi}\int dz\wedge d\bar{z}\sqrt{g}\left(g^{ab}\partial_{a}X^{1}\partial_{b}X^{1}-2iQRX^{1}\right)\,,\nonumber \\
 &  & \sigma=\ln g_{z\bar{z}}-\ln g_{z\bar{z}}^{\mathrm{\mathrm{A}}}\,,\nonumber \\
 &  & \Gamma\left[\sigma;g_{ab}\right]=-\frac{1}{4\pi}\int dz\wedge d\bar{z}\sqrt{g}\left(g^{ab}\partial_{a}\sigma\partial_{b}\sigma+2R\sigma\right)\,.\label{eq:Gamma}
\end{eqnarray}
$G^{\mathrm{A}}(z,w)$ denotes the Arakelov Green's function for the
Arakelov metric. The definitions of $g_{z\bar{z}}^{\mathrm{A}}$ and
$G^{\mathrm{A}}(z,w)$ are summarized in appendix \ref{sec:Arakelov-metric-and}.
The anomaly factor $e^{-\frac{1-12Q^{2}}{24}\Gamma\left[\sigma;\hat{g}_{z\bar{z}}\right]}$
is exactly what we expect for a theory with the central charge 
\begin{equation}
c=1-12Q^{2}
\end{equation}
of the linear dilaton conformal field theory.

$e^{ip\tilde{X}^{1}}$ thus defined turns out to be a primary field
with conformal dimension 
\begin{equation}
\frac{1}{2}p^{2}+Qp=\frac{1}{2}(p+Q)^{2}-\frac{Q^{2}}{2}\,.\label{eq:lineardilatondim}
\end{equation}
Notice that $\tilde{X}^{1}$ satisfies 
\begin{equation}
\partial\bar{\partial}\tilde{X}^{1}=0\,,
\end{equation}
if there are no source terms, and thus $i\partial\tilde{X}^{1}(z)$,
$i\bar{\partial}\tilde{X}^{1}(\bar{z})$ can be expanded as 
\begin{eqnarray}
i\partial\tilde{X}^{1}(z) & = & \sum_{n}\alpha_{n}^{1}z^{-n-1}\,,\nonumber \\
i\bar{\partial}\tilde{X}^{1}(\bar{z}) & = & \sum_{n}\bar{\alpha}_{n}^{1}\bar{z}^{-n-1}\,,
\end{eqnarray}
where $\alpha_{n}^{1}$ and $\bar{\alpha}_{n}^{1}$ satisfy the canonical
commutation relations. The states in the CFT are given as linear combinations
of the Fock space states 
\begin{equation}
\alpha_{-n_{1}}^{1}\cdots\alpha_{-n_{k}}^{1}\bar{\alpha}_{-\bar{n}_{1}}^{1}\cdots\bar{\alpha}_{-\bar{n}_{l}}^{1}\left|p\right\rangle \,,\label{eq:linearfock}
\end{equation}
where $|p\rangle=e^{ip\tilde{{X}^{1}}}(0)|0\rangle$. The states and
the oscillators satisfy
\begin{eqnarray}
\langle p_{1}|p_{2}\rangle & = & 2\pi\delta(p_{1}+p_{2}+2Q)\,,\nonumber \\
\left(\alpha_{n}^{1}\right)^{\ast} & = & -(\alpha_{-n}^{1}+2Q\delta_{n,0})\,,\nonumber \\
\left(\bar{\alpha}_{n}^{1}\right)^{\ast} & = & -(\bar{\alpha}_{-n}^{1}+2Q\delta_{n,0})\,,
\end{eqnarray}
where $\left\langle p\right|$, $\left(\alpha_{n}^{1}\right)^{\ast}$,
$\left(\bar{\alpha}_{n}^{1}\right)^{\ast}$ are the BPZ conjugates
of $\left|p\right\rangle $, $\alpha_{n}^{1},\bar{\alpha}_{n}^{1}$
respectively. On the sphere, the correlation function is given by
using the worldsheet metric $ds^{2}=dzd\bar{z}$ on the complex plane
as 
\begin{eqnarray}
 &  & \int\left[dX^{1}\right]_{g_{z\bar{z}}}e^{-S\left[X^{1};g_{z\bar{z}}\right]}\prod_{r=1}^{N}e^{ip_{r}\tilde{X}^{1}}(Z_{r},\bar{Z}_{r})\nonumber \\
 &  & \quad=2\pi\delta\left(\sum_{r}p_{r}+2Q\right)e^{-\frac{1-12Q^{2}}{24}\Gamma\left[\sigma;\frac{1}{2}\right]}\prod_{r>s}\left|Z_{r}-Z_{s}\right|^{2p_{r}p_{s}}\,.\label{eq:lincorrtree}
\end{eqnarray}
Using these, it is straightforward to construct the light-cone gauge
superstring field theory action in the background.

\subsection{Light-cone gauge superstring field theory in linear dilaton background}

Let us construct the light-cone gauge superstring field theory based
on the worldsheet theory with the variables 
\[
X^{i}~,\ \psi^{i}~,\ \bar{\psi}^{i}\quad(i=1,\cdots,8)\,,
\]
where the action for $X^{1}$, $\psi^{1}$, $\bar{\psi}^{1}$ is taken
to be (\ref{eq:linaction}) and that for other variables is the free
one. The worldsheet theory of the transverse variables turns out to
be a superconformal field theory with central charge 
\begin{equation}
c=12-12Q^{2}\,.
\end{equation}

The string field 
\[
\left|\Phi\left(t,\alpha\right)\right\rangle 
\]
is taken to be an element of the Hilbert space of the transverse variables
on the worldsheet and a function of 
\begin{eqnarray}
t & = & x^{+}\,,\nonumber \\
\alpha & = & 2p^{+}\,.
\end{eqnarray}
$\left|\Phi(t,\alpha)\right\rangle $ should be GSO even and satisfy
the level-matching condition 
\begin{equation}
(L_{0}-\bar{L}_{0})\left|\Phi\left(t,\alpha\right)\right\rangle =0\,,\label{eq:levelmatching}
\end{equation}
where $L_{0},\bar{L}_{0}$ are the zero modes of the Virasoro generators
of the worldsheet theory.

\begin{figure}
\begin{centering}
\includegraphics[scale=0.8]{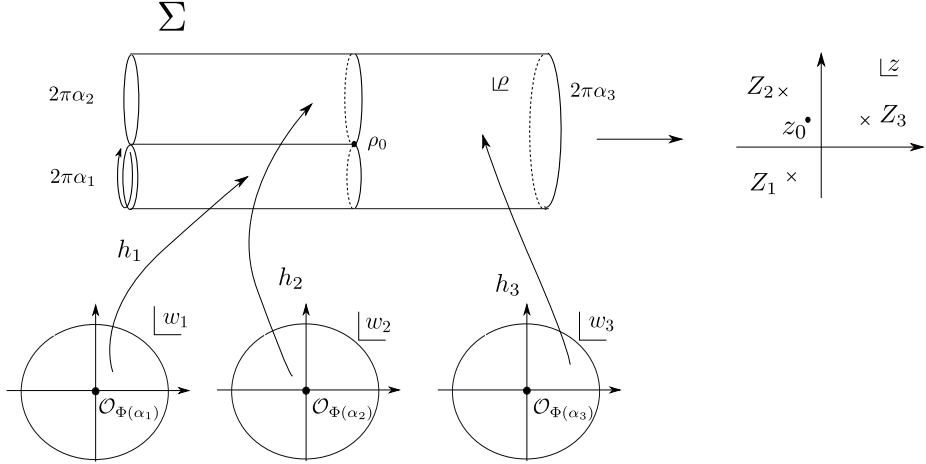} 
\par\end{centering}
\caption{The three string vertex for superstrings. Here we consider the case
$\alpha_{1},\alpha_{2}>0$, $\alpha_{3}<0$.\label{fig:The-three-string} }
\end{figure}

The action of the string field theory is given by \cite{Baba:2009kr,Ishibashi:2010nq}
\begin{eqnarray}
S & = & \int dt\left[\frac{1}{2}\sum_{\mathrm{B}}\int_{-\infty}^{\infty}\frac{\alpha d\alpha}{4\pi}\left\langle \Phi_{\mathrm{B}}\left(-\alpha\right)\right|(i\partial_{t}-\frac{L_{0}+\bar{L}_{0}-1+Q^{2}-i\varepsilon}{\alpha})\left|\Phi_{\mathrm{B}}\left(\alpha\right)\right\rangle \right.\nonumber \\
 &  & \ +\frac{1}{2}\sum_{\mathrm{F}}\int_{-\infty}^{\infty}\frac{d\alpha}{4\pi}\left\langle \Phi_{\mathrm{F}}\left(-\alpha\right)\right|(i\partial_{t}-\frac{L_{0}+\bar{L}_{0}-1+Q^{2}-i\varepsilon}{\alpha})\left|\Phi_{\mathrm{F}}\left(\alpha\right)\right\rangle \nonumber \\
 &  & \ -\frac{g_{s}}{6}\sum_{\mathrm{B}_{1},\mathrm{B}_{2},\mathrm{B}_{3}}\int\prod_{r=1}^{3}\left(\frac{\alpha_{r}d\alpha_{r}}{4\pi}\right)\delta\left(\sum_{r=1}^{3}\alpha_{r}\right)\left\langle V_{3}\left|\Phi_{\mathrm{B}_{1}}(\alpha_{1})\right.\right\rangle \left|\Phi_{\mathrm{B}_{2}}(\alpha_{2})\right\rangle \left|\Phi_{\mathrm{B}_{3}}(\alpha_{3})\right\rangle \nonumber \\
 &  & \ \left.-\frac{g_{s}}{2}\sum_{\mathrm{B}_{1},\mathrm{F}_{2},\mathrm{F}_{3}}\int\prod_{r=1}^{3}\left(\frac{\alpha_{r}d\alpha_{r}}{4\pi}\right)\delta\left(\sum_{r=1}^{3}\alpha_{r}\right)\left\langle V_{3}\left|\Phi_{\mathrm{B}_{1}}(\alpha_{1})\right.\right\rangle \alpha_{2}^{-\frac{1}{2}}\left|\Phi_{\mathrm{F}_{2}}(\alpha_{2})\right\rangle \alpha_{3}^{-\frac{1}{2}}\left|\Phi_{\mathrm{F}_{3}}(\alpha_{3})\right\rangle \right].\nonumber \\
 &  & \ \label{eq:superHamiltonian}
\end{eqnarray}
The first and the second terms are the kinetic terms with the Feynman
$i\varepsilon$ and $\left\langle \Phi(-\alpha)\right|$ denotes the
BPZ conjugate of $\left|\Phi(-\alpha)\right\rangle $. The third and
the fourth terms are the three string vertices and $g_{s}$ is the
string coupling constant. $\sum_{\mathrm{B}}$ and $\sum_{\mathrm{F}}$
denote the sums over bosonic and fermionic string fields respectively.
By the state-operator correspondence of the worldsheet conformal field
theory, there exists a local operator $\mathcal{O}_{\Phi}(w,\bar{w})$
corresponding to any state $\left|\Phi\right\rangle $. $\left\langle \left.V_{3}\right|\Phi(\alpha_{1})\right\rangle \left|\Phi(\alpha_{2})\right\rangle \left|\Phi(\alpha_{3})\right\rangle $
with $\sum_{r=1}^{3}\alpha_{r}=0$ is defined to be 
\begin{eqnarray}
\lefteqn{\left\langle \left.V_{3}\right|\Phi(\alpha_{1})\right\rangle \left|\Phi(\alpha_{2})\right\rangle \left|\Phi(\alpha_{3})\right\rangle }\nonumber \\
 & = & \left\langle \lim_{\rho\to\rho_{0}}\left|\rho-\rho_{0}\right|^{\frac{3}{2}}T_{F}^{LC}\left(\rho\right)\bar{T}_{F}^{LC}\left(\bar{\rho}\right)h_{1}\circ\mathcal{O}_{\Phi\left(\alpha_{1}\right)}(0,0)h_{2}\circ\mathcal{O}_{\Phi\left(\alpha_{2}\right)}(0,0)h_{3}\circ\mathcal{O}_{\Phi\left(\alpha_{3}\right)}(0,0)\right\rangle _{\Sigma}\,,\nonumber \\
\label{eq:bosonicthree}
\end{eqnarray}
in terms of a correlation function on $\Sigma$, which is the worldsheet
describing the three string interaction depicted in figure \ref{fig:The-three-string}.
On each cylinder corresponding to an external line, one can introduce
a complex coordinate 
\begin{equation}
\rho=\tau+i\sigma\,,\label{eq:rho-coordinate}
\end{equation}
whose real part $\tau$ coincides with the Wick rotated light-cone
time $it$ and imaginary part $\sigma\sim\sigma+2\pi\alpha_{r}$ parametrizes
the closed string at each time. The $\rho$'s on the cylinders are
smoothly connected except at the interaction point $\rho_{0}$ and
we get a complex coordinate $\rho$ on $\Sigma$. The correlation
function $\left\langle \quad\right\rangle _{\Sigma}$ is defined with
the metric 
\begin{equation}
ds^{2}=d\rho d\bar{\rho}\,,\label{eq:canonical-metric}
\end{equation}
on the worldsheet. $h_{r}(w_{r})$ gives a map from a unit disk $\left|w_{r}\right|<1$
to the cylinder corresponding to the $r$-th external line so that
\begin{equation}
w_{r}=e^{\frac{1}{\alpha_{r}}(h_{r}(w_{r})-\rho_{0})}\,.
\end{equation}
$T_{F}^{LC}$, $\bar{T}_{F}^{LC}$ are the supercurrents of the transverse
worldsheet theory.

\subsection{Feynman amplitudes of light-cone gauge superstring field theory in
linear dilaton background }

\begin{figure}
\begin{centering}
\includegraphics[scale=0.5]{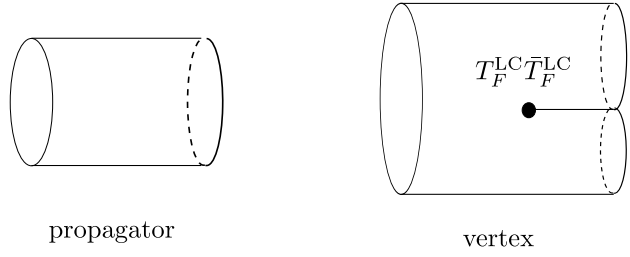} 
\par\end{centering}
\caption{The propagator and the vertex of the string field theory. \label{fig:The-propagator-and}}
\end{figure}

\begin{figure}
\begin{centering}
\includegraphics[scale=1.4]{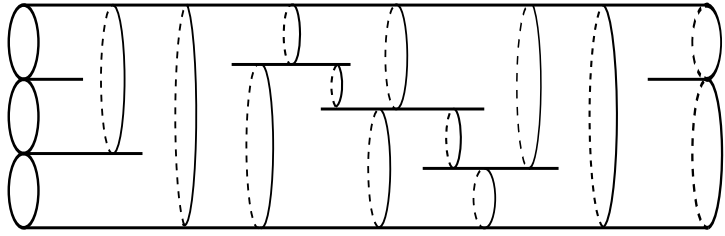} 
\par\end{centering}
\caption{A string diagram with $3$ incoming, $2$ outgoing strings and $3$
loops. \label{fig:A-string-diagram}}
\end{figure}

It is straightforward to calculate the amplitudes by the old-fashioned
perturbation theory starting from the action (\ref{eq:superHamiltonian})
and Wick rotate to Euclidean time. The propagator and the vertex are
given by the worldsheets depicted in figure \ref{fig:The-propagator-and}.
Each term in the expansion corresponds to a light-cone gauge Feynman
diagram for strings. A typical diagram is depicted in figure \ref{fig:A-string-diagram}.

A Wick rotated $g$-loop $N$-string diagram is conformally equivalent
to an $N$ punctured genus $g$ Riemann surface $\Sigma$. A $g$-loop
$N$-string amplitude is given as an integral over the moduli space
of $\Sigma$ as \cite{D'Hoker:1987pr,Aoki:1990yn}
\begin{equation}
\mathcal{A}_{N}^{(g)}=(ig_{s})^{2g-2+N}C\int[dT][\alpha d\theta][d\alpha]\,F_{N}^{(g)}~,\label{eq:ANg}
\end{equation}
where $\int[dT][\alpha d\theta][d\alpha]$ denotes the integration
over the moduli parameters and $C$ is the combinatorial factor. In
each channel, the integration measure is given as 
\begin{equation}
\int[dT][\alpha d\theta][d\alpha]=\prod_{a=1}^{2g-3+N}\left(-i\int_{0}^{\infty}dT_{a}\right)\prod_{A=1}^{g}\int\frac{d\alpha_{A}}{4\pi}\prod_{\mathcal{I}=1}^{3g-3+N}\left(|\alpha_{\mathcal{I}}|\int_{0}^{2\pi}\frac{d\theta_{\mathcal{I}}}{2\pi}\right).
\end{equation}
Here $T_{a}$'s are heights of the cylinders corresponding to internal
lines, $\alpha_{A}$'s denote the circumferences of the cylinders
corresponding to the $+$ components of the loop momenta and $\alpha_{\mathcal{I}}$'s
and $\theta_{\mathcal{I}}$'s are the string-lengths and the twist
angles for the internal propagators.

The integrand $F_{N}^{(g)}$ is given as a path integral over the
transverse variables $X^{i}$, $\psi^{i}$, $\bar{\psi}^{i}$ $\left(i=1,\ldots,8\right)$
on the light-cone diagram. A light-cone diagram consists of cylinders
which correspond to propagators of the closed string as mentioned
above. On each cylinder, one can introduce a complex coordinate $\rho$
in the same way as the complex coordinate (\ref{eq:rho-coordinate})
is introduced on the three string vertex. The $\rho$'s on the cylinders
are smoothly connected except at the interaction points and we get
a complex coordinate $\rho$ on the light-cone diagram $\Sigma$.
The path integral on the light-cone diagram is defined by using the
metric (\ref{eq:canonical-metric}).

$\rho$ is not a good coordinate around the interaction points and
the punctures, and the metric (\ref{eq:canonical-metric}) is not
well-defined at these points. $F_{N}^{(g)}$ can be expressed in terms
of correlation functions defined with a metric $d\hat{s}^{2}=2\hat{g}_{z\bar{z}}dzd\bar{z}$
which is regular everywhere on the worldsheet, as 
\begin{eqnarray}
F_{N}^{(g)} & = & \left(2\pi\right)^{2}\delta\left(\sum_{r=1}^{N}p_{r}^{+}\right)\delta\left(\sum_{r=1}^{N}p_{r}^{-}\right)e^{-\frac{1}{2}(1-Q^{2})\Gamma\left[\sigma;\hat{g}_{z\bar{z}}\right]}\nonumber \\
 &  & \times\int\left[dX^{i}d\psi^{i}d\bar{\psi}^{i}\right]_{\hat{g}_{z\bar{z}}}e^{-S^{\mathrm{LC}}\left[X^{i},\psi^{i},\bar{\psi}^{i}\right]}\prod_{I=1}^{2g-2+N}\left(\left|\partial^{2}\rho\left(z_{I}\right)\right|^{-\frac{3}{2}}T_{F}^{\mathrm{LC}}\left(z_{I}\right)\bar{T}_{F}^{\mathrm{LC}}\left(\bar{z}_{I}\right)\right)\prod_{r=1}^{N}V_{r}^{\mathrm{LC}}\,.\nonumber \\
\label{eq:superFN}
\end{eqnarray}
Here $S^{\mathrm{LC}}\left[X^{i},\psi^{i},\bar{\psi}^{i}\right]$
denotes the worldsheet action of the transverse variables and the
path integral measure $\left[dX^{i}d\psi^{i}d\bar{\psi}^{i}\right]_{\hat{g}_{z\bar{z}}}$
is defined with the metric $d\hat{s}^{2}=2\hat{g}_{z\bar{z}}dzd\bar{z}$.
Since the integrand should be defined by using the canonical metric
$ds^{2}=\partial\rho\bar{\partial}\bar{\rho}dzd\bar{z}$ on the light-cone
diagram, we need the anomaly factor $e^{-\frac{1}{2}(1-Q^{2})\Gamma\left[\sigma;\hat{g}_{z\bar{z}}\right]}$,
where 
\begin{eqnarray}
 &  & \sigma=\ln\partial\rho\bar{\partial}\bar{\rho}-\ln\hat{g}_{z\bar{z}}\,,\nonumber \\
 &  & \Gamma\left[\sigma;\hat{g}_{z\bar{z}}\right]=-\frac{1}{4\pi}\int dz\wedge d\bar{z}\sqrt{\hat{g}}\left(\hat{g}^{ab}\partial_{a}\sigma\partial_{b}\sigma+2\hat{R}\sigma\right)\,.\label{eq:Gammasigma}
\end{eqnarray}
$V_{r}^{\mathrm{LC}}$ denotes the vertex operator for the $r$-th
external line. When the $r$-th external line corresponds to the state
\begin{equation}
\alpha_{-n_{1}}^{i_{1}}\cdots\bar{\alpha}_{-\bar{n}_{1}}^{\bar{i}_{1}}\cdots\psi_{-s_{1}}^{j_{1}}\cdots\bar{\psi}_{-\bar{s}_{1}}^{\bar{j}_{1}}\cdots\left|p_{r}\right\rangle 
\end{equation}
in the (NS,NS) sector, the light-cone vertex $V_{r}^{\mathrm{LC}}$
is given as 
\begin{eqnarray}
V_{r}^{\mathrm{LC}} & = & \alpha_{r}\oint_{0}\frac{dw_{r}}{2\pi i}i\partial\tilde{X}^{i_{1}}(w_{r})w_{r}^{-n_{1}}\cdots\oint_{0}\frac{d\bar{w}_{r}}{2\pi i}i\bar{\partial}\tilde{X}^{\bar{i}_{1}}(\bar{w}_{r})\bar{w}_{r}^{-\bar{n}_{1}}\cdots\nonumber \\
 &  & \quad\times\oint_{0}\frac{dw_{r}}{2\pi i}\psi^{j_{1}}(w_{r})w_{r}^{-s_{1}-\frac{1}{2}}\cdots\oint_{0}\frac{d\bar{w}_{r}}{2\pi i}\psi^{\bar{j}_{1}}(\bar{w}_{r})\bar{w}_{r}^{-\bar{s}_{1}-\frac{1}{2}}\cdots\nonumber \\
 &  & \quad\times e^{i\vec{p}_{r}\cdot\vec{\tilde{X}}}\left(w_{r}=0,\bar{w}_{r}=0\right)e^{-p_{r}^{-}\tau_{0}^{\left(r\right)}}\,.\label{eq:superVLC}
\end{eqnarray}
Here 
\begin{eqnarray}
\tilde{X}^{i} & \equiv & X^{i}-iQ\delta^{i1}\ln(2g_{z\bar{z}})\,,\nonumber \\
w_{r} & \equiv & \exp\left[\frac{1}{\alpha_{r}}\left(\rho\left(z\right)-\rho(z_{I^{(r)}})\right)\right]\,,\label{eq:wr}\\
\tau_{0}^{(r)} & \equiv & \mathop{\mathrm{Re}}\rho\left(z_{I^{\left(r\right)}}\right)\,,\label{eq:tau0r}
\end{eqnarray}
and $z_{I^{\left(r\right)}}$ is defined to be the coordinate of the
interaction point at which the $r$-th external line interacts. The
on-shell and the level-matching conditions are 
\begin{equation}
\frac{1}{2}\left(-2p_{r}^{+}p_{r}^{-}+p_{r}^{i}p_{r}^{i}\right)+Qp_{r}^{1}+\mathcal{N}_{r}=\frac{1}{2}(1-Q^{2})~,\quad\mathcal{N}_{r}\equiv\sum_{k}n_{k}+\sum_{l}s_{l}=\sum_{\bar{k}}\bar{n}_{\bar{k}}+\sum_{\bar{l}}\bar{s}_{\bar{l}}~.\label{eq:superonshell}
\end{equation}

It is possible to calculate the right hand side of (\ref{eq:superFN}).
$\rho$ can be given as a function of local coordinate $z$ on $\Sigma$
as 
\begin{equation}
\rho(z)=\sum_{r=1}^{N}\alpha_{r}\left[\ln E(z,Z_{r})-2\pi i\int_{P_{0}}^{z}\omega\frac{1}{\mathrm{Im}\Omega}\mathrm{Im}\int_{P_{0}}^{Z_{r}}\omega\right]~,\qquad\sum_{r=1}^{N}\alpha_{r}=0~,\label{eq:mandelstammulti}
\end{equation}
up to an additive constant independent of $z$. Here $E(z,w)$ is
the prime form of the surface, $\omega$ is the canonical basis of
the holomorphic abelian differentials and $\Omega$ is the period
matrix.\footnote{For the mathematical background relevant for string perturbation theory,
we refer the reader to \cite{D'Hoker:1988ta}.} The base point $P_{0}$ is arbitrary. There are $2g-2+N$ zeros of
$\partial\rho$ and we denote them by $z_{I}\,(I=1,\cdots,2g-2+N)$.
They correspond to the interaction points of the light-cone diagram.
Substituting (\ref{eq:mandelstammulti}) into (\ref{eq:Gammasigma})
yields a divergent result for $\Gamma\left[\sigma;\hat{g}_{z\bar{z}}^{\mathrm{}}\right]$.
We can obtain $e^{-\Gamma\left[\sigma;\hat{g}_{z\bar{z}}^{\mathrm{}}\right]}$
up to a divergent numerical factor by regularizing it as was done
in \cite{Mandelstam:1985ww}. The divergent factor can be absorbed
in a redefinition of $g_{s}$ and the vertex operator. Taking $\hat{g}_{z\bar{z}}$
to be the Arakelov metric \cite{arakelov}, $e^{-\Gamma\left[\sigma;g_{z\bar{z}}^{\mathrm{A}}\right]}$
for higher genus surfaces is calculated in \cite{Ishibashi:2013nma}
to be 
\begin{equation}
e^{-\Gamma\left[\sigma;g_{z\bar{z}}^{\mathrm{A}}\right]}\propto e^{-W}\prod_{r}e^{-2\mathop{\mathrm{Re}}\bar{N}_{00}^{rr}}\prod_{I}\left|\partial^{2}\rho\left(z_{I}\right)\right|^{-3}\,,\label{eq:e^-Gamma}
\end{equation}
up to a numerical constant which can be fixed by imposing the factorization
condition. Here 
\begin{eqnarray}
-W & \equiv & -2\sum_{I<J}G^{\mathrm{A}}\left(z_{I};z_{J}\right)-2\sum_{r<s}G^{\mathrm{A}}\left(Z_{r};Z_{s}\right)+2\sum_{I,r}G^{\mathrm{A}}\left(z_{I};Z_{r}\right)\nonumber \\
 &  & {}-\sum_{r}\ln\left(2g_{Z_{r}\bar{Z}_{r}}^{\mathrm{A}}\right)+3\sum_{I}\ln\left(2g_{z_{I}\bar{z}_{I}}^{\mathrm{A}}\right)\, ,\nonumber \\
\bar{N}_{00}^{rr} & \equiv & \lim_{z\to Z_{r}}\left[\frac{\rho(z_{I^{(r)}})-\rho(z)}{\alpha_{r}}+\ln(z-Z_{r})\right]\nonumber \\
 & = & \frac{\rho(z_{I^{(r)}})}{\alpha_{r}}-\sum_{s\neq r}\frac{\alpha_{s}}{\alpha_{r}}\ln E(Z_{r},Z_{s})+\frac{2\pi i}{\alpha_{r}}\int_{P_{0}}^{Z_{r}}\omega\frac{1}{\mathop{\mathrm{Im}}\Omega}\sum_{s=1}^{N}\alpha_{s}\mathop{\mathrm{Im}}\int_{P_{0}}^{Z_{s}}\omega~.
\end{eqnarray}
The correlation functions of $X^{i},\psi^{i},\bar{\psi}^{i}$ which
appear in (\ref{eq:superFN}) can be calculated by using the formulas
given in \cite{AlvarezGaume:1987vm,Verlinde:1986kw,Dugan:1987qe,Sonoda1987c,Atick1987b}
and (\ref{eq:linearcorr}). From the explicit form of the integrand
$F_{N}^{(g)}$, one can see that the amplitude $A_{N}^{(g)}$ suffers
from divergences due to the collisions of the interaction points and
the degenerations of the surface $\Sigma$, if $Q=0$. In \cite{Ishibashi2016a},
it was shown that $A_{N}^{(g)}$ becomes finite if one takes $Q^{2}>10$
and $\varepsilon>0$. Therefore it is possible to define $A_{N}^{(g)}$
as an analytic function of $Q$ and take the limit $Q\to0$ as is
usually done in dimensional regularization of field theory.

\section{BRST invariant form of the amplitudes\label{sec:BRST-invariant-form}}

We would like to show that the integrand (\ref{eq:superFN}) can be
rewritten in terms of a correlation function of the conformal gauge
variables $X^{\mu},\psi^{\mu},\bar{\psi}^{\mu}$ ($\mu=+,-,1,\ldots,8$),
$b,c,\bar{b},\bar{c}$, $\beta,\gamma,\bar{\beta},\bar{\gamma}$.
For $Q\ne0$, the worldsheet theory of the longitudinal variables
$X^{\pm},\psi^{\pm},\bar{\psi}^{\pm}$ becomes so-called $X^{\pm}$ CFT, which is defined
and analyzed in \cite{Baba:2009fi,Ishibashi2016b}. The results of
these references are summarized in appendix \ref{sec:Supersymmetric--CFT}.
Using (\ref{eq:superXpmcorr2}) and (\ref{eq:Gammasupermulti2}) there,
one can prove 
\begin{eqnarray}
 &  & (2\pi)^{2}\delta\left(\sum_{r=1}^{N}p_{r}^{+}\right)\delta\left(\sum_{r=1}^{N}p_{r}^{-}\right)\left(Z^{X}[\hat{g}_{z\bar{z}}]Z^{\psi}[\hat{g}_{z\bar{z}}]\right)^{2}\prod_{r=1}^{N}V_{r}^{\mathrm{LC}}\nonumber \\
 &  & =\prod_{r=1}^{N}\left(\alpha_{r}e^{\mathop{\mathrm{Re}}\bar{N}_{00}^{rr}}\right)e^{-\frac{Q^{2}}{2}\Gamma\left[\sigma;\hat{g}_{z\bar{z}}\right]}\nonumber \\
 &  & \times\int\left[d\mathcal{X}^{+}d\mathcal{X}^{-}\right]_{\hat{g}_{z\bar{z}}}e^{-S_{\mathrm{super}}^{\pm}\left[\mathcal{X}^{\pm};\hat{g}_{z\bar{z}}\right]}\prod_{r=1}^{N}\left[\oint_{z_{I^{\left(r\right)}}}\frac{d\mathbf{z}}{2\pi i}\mathcal{S}\left(\mathbf{z},Z_{r}\right)\oint_{\bar{z}_{I^{\left(r\right)}}}\frac{d\bar{\mathbf{z}}}{2\pi i}\bar{\mathcal{S}}\left(\bar{\mathbf{z}},\bar{Z}_{r}\right)V_{r}^{\mathrm{DDF}}(Z_{r},\bar{Z}_{r})\right].\nonumber \\
 &  & \ \label{eq:superVLC-DDF}
\end{eqnarray}
Here $\mathcal{X}^{\pm}$ is the superfield given in (\ref{eq:superfieldXpm}).
The supersymmetric contour integral is defined as 
\begin{equation}
\oint_{z_{I^{\left(r\right)}}}\frac{d\mathbf{z}}{2\pi i}=\oint_{z_{I^{\left(r\right)}}}\frac{dz}{2\pi i}\int d\theta\,,
\end{equation}
using the Berezinian integral $\int d\theta$, and the antiholomorphic
version is defined similarly. $\mathcal{S}\left(\mathbf{z},w\right)$
is defined to be 
\begin{equation}
\mathcal{S}(\mathbf{\mathbf{z}},w)\equiv D\ln\left(\partial\mathcal{X}^{+}-\frac{\partial D\mathcal{X}^{+}D\mathcal{X}^{+}}{\partial\mathcal{X}^{+}}\right)\left(\mathbf{z}\right)e^{-i\frac{Q^{2}}{\alpha_{r}}\left(\mathcal{X}_{L}^{+}\left(\mathbf{z}\right)-X_{L}^{+}\left(w\right)\right)}\,,
\end{equation}
and $\bar{\mathcal{S}}\left(\bar{\mathbf{z}},w\right)$ is the antiholomorphic
version. $V_{r}^{\mathrm{DDF}}(Z_{r},\bar{Z}_{r})$ is the supersymmetric
DDF vertex operator given by 
\begin{equation}
V_{r}^{\mathrm{DDF}}(Z_{r},\bar{Z}_{r})=A_{-n_{1}}^{i_{1}(r)}\cdots\bar{A}_{-\bar{n}_{1}}^{\bar{i}_{1}(r)}\cdots B_{-s_{1}}^{j_{1}(r)}\cdots\bar{B}_{-\bar{s}_{1}}^{\bar{j}_{1}(r)}\cdots e^{-ip_{r}^{+}X^{-}-i\left(p_{r}^{-}-\frac{\mathcal{N}_{r}}{p_{r}^{+}}\right)X^{+}+ip_{r}^{i}\tilde{X}^{i}}(Z_{r},\bar{Z}_{r})~,\label{eq:superVDDF}
\end{equation}
with the DDF operators $A_{-n}^{i(r)},B_{-s}^{j(r)}$ for the $r$-th
string defined as 
\begin{eqnarray}
A_{-n}^{i(r)} & = & \oint_{Z_{r}}\frac{d\mathbf{z}}{2\pi i}iD\left(\tilde{\mathcal{X}}^{i}+iQ\delta^{i,1}\Phi\right)e^{-i\frac{n}{p_{r}^{+}}\mathcal{X}_{L}^{+}}(\mathbf{z})~,\nonumber \\
B_{-s}^{i(r)} & = & \oint_{Z_{r}}\frac{d\mathbf{z}}{2\pi i}\frac{1}{\left(ip_{r}^{+}\right)^{\frac{1}{2}}}\Theta^{+}D\left(\mathcal{\tilde{X}}^{i}+iQ\delta^{i,1}\Phi\right)e^{-i\frac{s}{p_{r}^{+}}\mathcal{X}_{L}^{+}}(\mathbf{z})~,
\end{eqnarray}
where $\tilde{\mathcal{X}}^{i}$ is the superfield for $\tilde{X}^{i},\psi^{i},\bar{\psi}^{i}$,
$\Phi$ is defined in (\ref{eq:Phi}) and $\mathcal{X}_{L}^{+}$ denotes
the left moving part of $\mathcal{X}^{+}$. $\bar{A}_{-n}^{i(r)},\bar{B}_{-s}^{i(r)}$
are similarly given for the antiholomorphic sector. The product $\mathcal{S}\bar{\mathcal{S}}V_{r}^{\mathrm{DDF}}$
is normal ordered as 
\begin{eqnarray}
 &  & \mathcal{S}\left(\mathbf{z},Z_{r}\right)\bar{\mathcal{S}}\left(\bar{\mathbf{z}},\bar{Z}_{r}\right)V_{r}^{\mathrm{DDF}}\left(Z_{r},\bar{Z}_{r}\right)\nonumber \\
 &  & \quad\equiv\lim_{w\to Z_{r}}\lim_{\bar{w}\to\bar{Z}_{r}}\mathcal{S}\left(\mathbf{z},w\right)\bar{\mathcal{S}}\left(\bar{\mathbf{z}},\bar{w}\right)V_{r}^{\mathrm{DDF}}\left(Z_{r},\bar{Z}_{r}\right)\left|w-Z_{r}\right|^{-Q^{2}}\,.
\end{eqnarray}
Substituting (\ref{eq:superVLC-DDF}) into (\ref{eq:superFN}), we
find that the integrand $F_{N}^{(g)}$ can be expressed as 
\begin{eqnarray}
F_{N}^{(g)} & \propto & \prod_{r=1}^{N}\left(\alpha_{r}e^{\mathop{\mathrm{Re}}\bar{N}_{00}^{rr}}\right)\prod_{I}\left|\partial^{2}\rho\left(z_{I}\right)\right|^{-\frac{3}{2}}e^{-\frac{1}{2}\Gamma\left[g_{z\bar{z}}^{\mathrm{A}},\,\ln|\partial\rho|^{2}\right]}\left(Z^{X}[g_{z\bar{z}}^{\mathrm{A}}]Z^{\psi}[g_{z\bar{z}}^{\mathrm{A}}]\right)^{-2}\nonumber \\
 &  & \times\int\left[d\mathcal{X}^{\mu}\right]_{\hat{g}_{z\bar{z}}}e^{-S_{\mathrm{super}}^{\pm}\left[\mathcal{X}^{\pm};\hat{g}_{z\bar{z}}\right]-S^{\mathrm{LC}}\left[X^{i},\psi^{i},\bar{\psi}^{i}\right]}\prod_{I}\left[T_{F}^{\mathrm{LC}}\left(z_{I}\right)\bar{T}_{F}^{\mathrm{LC}}\left(\bar{z}_{I}\right)\right]\nonumber \\
 &  & \hphantom{\int\left[d\mathcal{X}^{\mu}d\mathcal{X}^{\mu}\right]}\times\prod_{r=1}^{N}\left[\oint_{z_{I^{\left(r\right)}}}\frac{d\mathbf{z}}{2\pi i}\mathcal{S}\left(\mathbf{z},Z_{r}\right)\oint_{\bar{z}_{I^{\left(r\right)}}}\frac{d\bar{\mathbf{z}}}{2\pi i}\bar{\mathcal{S}}\left(\bar{\mathbf{z}},\bar{Z}_{r}\right)V_{r}^{\mathrm{DDF}}(Z_{r},\bar{Z}_{r})\right].~~~\label{eq:superFN2}
\end{eqnarray}

We will further rewrite (\ref{eq:superFN2}) by introducing the ghost
variables. The identities satisfied by the ghost correlation functions
which should be used here are summarized in appendix \ref{sec:-systems-on-higher}.
Taking the metric $\hat{g}_{z\bar{z}}$ to be the Arakelov metric
$g_{z\bar{z}}^{\mathrm{A}}$ and using (\ref{eq:superFN2}), (\ref{eq:superghostZpsi})
and (\ref{eq:ZXcontour}), the amplitude (\ref{eq:ANg}) can be rewritten
as 
\begin{eqnarray}
A_{N}^{(g)} & \propto & \int[dT][d\alpha][\alpha d\theta]\nonumber \\
 &  & \times\int\left[dX^{\mu}d\psi^{\mu}d\bar{\psi}^{\mu}dbd\bar{b}dcd\bar{c}d\beta d\bar{\beta}d\gamma d\bar{\gamma}\right]_{g_{z\bar{z}}^{\mathrm{A}}}e^{-S^{\mathrm{tot}}}\nonumber \\
 &  & \qquad\times\prod_{K=1}^{6g-6+2N}\left[\oint_{C_{K}}\frac{dz}{\partial\rho}b_{zz}+\varepsilon_{K}\oint_{\bar{C}_{K}}\frac{d\bar{z}}{\bar{\partial}\bar{\rho}}b_{\bar{z}\bar{z}}\right]\prod_{I}\left[e^{\phi}T_{F}^{\mathrm{LC}}\left(z_{I}\right)e^{\bar{\phi}}\bar{T}_{F}^{\mathrm{LC}}\left(\bar{z}_{I}\right)\right]\nonumber \\
 &  & \qquad\times\prod_{r=1}^{N}\left[\oint_{z_{I^{\left(r\right)}}}\frac{d\mathbf{z}}{2\pi i}\mathcal{S}\left(\mathbf{z},Z_{r}\right)\oint_{\bar{z}_{I^{\left(r\right)}}}\frac{d\bar{\mathbf{z}}}{2\pi i}\bar{\mathcal{S}}\left(\bar{\mathbf{z}},\bar{Z}_{r}\right)c\bar{c}e^{-\phi-\bar{\phi}}V_{r}^{\mathrm{DDF}}(Z_{r},\bar{Z}_{r})\right]\,.\label{eq:superFN3}
\end{eqnarray}
Here $S^{\mathrm{tot}}$ denotes the worldsheet action for the variables
$X^{\mu},\psi^{\mu},\bar{\psi}^{\mu}$ and the ghosts. It is shown
in \cite{Baba:2009fi} that the worldsheet theory becomes a conformal
field theory with vanishing central charge, and we can define the nilpotent
BRST operator $Q_{\mathrm{B}}$. The quantities which appear in this
expression would be BRST invariant, if $e^{\phi}T_{F}^{\mathrm{LC}}\left(z_{I}\right)$
and $e^{\bar{\phi}}\bar{T}_{F}^{\mathrm{LC}}\left(\bar{z}_{I}\right)$
were the PCO's. Actually (\ref{eq:superFN3}) can be turned into a
BRST invariant form as 
\begin{eqnarray}
A_{N}^{(g)} & \propto & \int[dT][d\alpha][\alpha d\theta]\nonumber \\
 &  & \times\int\left[dX^{\mu}d\psi^{\mu}d\bar{\psi}^{\mu}dbd\bar{b}dcd\bar{c}d\beta d\bar{\beta}d\gamma d\bar{\gamma}\right]_{g_{z\bar{z}}^{\mathrm{A}}}e^{-S^{\mathrm{tot}}}\nonumber \\
 &  & \qquad\times\prod_{K=1}^{6g-6+2N}\left[\oint_{C_{K}}\frac{dz}{\partial\rho}b_{zz}+\varepsilon_{K}\oint_{\bar{C}_{K}}\frac{d\bar{z}}{\bar{\partial}\bar{\rho}}b_{\bar{z}\bar{z}}\right]\prod_{I}\left[X\left(z_{I}\right)\bar{X}\left(\bar{z}_{I}\right)\right]\nonumber \\
 &  & \qquad\times\prod_{r=1}^{N}\left[\oint_{z_{I^{(r)}}}\frac{d\mathbf{z}}{2\pi i}\mathcal{S}\left(\mathbf{z},Z_{r}\right)\oint_{\bar{z}_{I^{(r)}}}\frac{d\bar{\mathbf{z}}}{2\pi i}\bar{\mathcal{S}}\left(\bar{\mathbf{z}},\bar{Z}_{r}\right)c\bar{c}e^{-\phi-\bar{\phi}}V_{r}^{\mathrm{DDF}}(Z_{r},\bar{Z}_{r})\right]\,,\label{eq:superBRSTinvariant}
\end{eqnarray}
where 
\begin{equation}
X\left(z\right)=\left[c\partial\xi-e^{\phi}T_{F}+\frac{1}{4}\partial b\eta e^{2\phi}+\frac{1}{4}b\left(2\partial\eta e^{2\phi}+\eta\partial e^{2\phi}\right)\right](z)\label{eq:PCO}
\end{equation}
is the PCO and $\bar{X}\left(\bar{z}\right)$ is its antiholomorphic
counterpart. Here $T_{F}$ denotes the supercurrent for $\mathcal{X}^{\mu}$
$(\mu=+,-,1,\ldots,8)$. Since the PCO's and the contour integrals
of $\mathcal{S}\left(\mathbf{z},Z_{r}\right),\bar{\mathcal{S}}\left(\bar{\mathbf{z}},\bar{Z}_{r}\right)$
do not commute, we need to be a little careful about the definition
of the right hand side of (\ref{eq:superBRSTinvariant}). To be precise,
the right hand side of (\ref{eq:superBRSTinvariant}) should be defined
as 
\begin{eqnarray}
 &  & \int[dT][d\alpha][\alpha d\theta]\nonumber \\
 &  & \times\int\left[dX^{\mu}d\psi^{\mu}dbd\bar{b}dcd\bar{c}d\beta d\bar{\beta}d\gamma d\bar{\gamma}\right]_{g_{z\bar{z}}^{\mathrm{A}}}e^{-S^{\mathrm{tot}}}\nonumber \\
 &  & \qquad\times\prod_{K=1}^{6g-6+2N}\left[\oint_{C_{K}}\frac{dz}{\partial\rho}b_{zz}+\varepsilon_{K}\oint_{\bar{C}_{K}}\frac{d\bar{z}}{\bar{\partial}\bar{\rho}}b_{\bar{z}\bar{z}}\right]\lim_{\epsilon\to0}\prod_{I}\left[X\left(z_{I}+2\epsilon\right)\bar{X}\left(\bar{z}_{I}+2\epsilon\right)\right]\nonumber \\
 &  & \qquad\times\prod_{r=1}^{N}\left[\oint_{C_{r,\left|\epsilon\right|}}\frac{d\mathbf{z}}{2\pi i}\mathcal{S}\left(\mathbf{z},Z_{r}\right)\oint_{\bar{C}_{r,\left|\epsilon\right|}}\frac{d\bar{\mathbf{z}}}{2\pi i}\bar{\mathcal{S}}\left(\bar{\mathbf{z}},\bar{Z}_{r}\right)c\bar{c}e^{-\phi-\bar{\phi}}V_{r}^{\mathrm{DDF}}(Z_{r},\bar{Z}_{r})\right]\,.\label{eq:superBRSTinvariant2}
\end{eqnarray}
The contour $C_{r,\left|\epsilon\right|}$ is a circle with radius
$\left|\epsilon\right|$ around $z=z_{I^{\left(r\right)}}$. One can
show 
\begin{eqnarray}
 &  & \left[Q_{\mathrm{B}},\oint_{C_{r,\left|\epsilon\right|}}\frac{d\mathbf{z}}{2\pi i}\mathcal{S}\left(\mathbf{z},Z_{r}\right)\oint_{\bar{C}_{r,\left|\epsilon\right|}}\frac{d\bar{\mathbf{z}}}{2\pi i}\bar{\mathcal{S}}\left(\bar{\mathbf{z}},\bar{Z}_{r}\right)c\bar{c}e^{-\phi-\bar{\phi}}V_{r}^{\mathrm{DDF}}(Z_{r},\bar{Z}_{r})\right]\nonumber \\
 &  & \quad=\oint_{C_{r,\left|\epsilon\right|}}\frac{d\mathbf{z}}{2\pi i}\partial DC\left(\mathbf{z}\right)e^{-i\frac{Q^{2}}{\alpha_{r}}\left(\mathcal{X}_{L}^{+}\left(\mathbf{z}\right)-X_{L}^{+}\left(Z_{r}\right)\right)}\oint_{\bar{C}_{r,\left|\epsilon\right|}}\frac{d\bar{\mathbf{z}}}{2\pi i}\bar{\mathcal{S}}\left(\bar{\mathbf{z}},\bar{Z}_{r}\right)c\bar{c}e^{-\phi-\bar{\phi}}V_{r}^{\mathrm{DDF}}(Z_{r},\bar{Z}_{r})\nonumber \\
 &  & \hphantom{\quad=}+\oint_{C_{r,\left|\epsilon\right|}}\frac{d\mathbf{z}}{2\pi i}\mathcal{S}\left(\mathbf{z},Z_{r}\right)\oint_{\bar{C}_{r,\left|\epsilon\right|}}\frac{d\bar{\mathbf{z}}}{2\pi i}\bar{\partial}\bar{D}C\left(\bar{\mathbf{z}}\right)e^{-i\frac{Q^{2}}{\alpha_{r}}\left(\mathcal{X}_{R}^{+}\left(\bar{\mathbf{z}}\right)-X_{R}^{+}\left(\bar{Z_{r}}\right)\right)}c\bar{c}e^{-\phi-\bar{\phi}}V_{r}^{\mathrm{DDF}}(Z_{r},\bar{Z}_{r})\,,\nonumber \\
\end{eqnarray}
and with no operators inside $C_{r,\left|\epsilon\right|},\bar{C}_{r,\left|\epsilon\right|}$,
this vanishes. All the other factors are BRST invariant in the usual
way and thus one can prove that (\ref{eq:superBRSTinvariant}) is
a BRST invariant expression. When there exists no $r\,\left(r=1,\ldots,N\right)$
such that $z_{I}=z_{I^{\left(r\right)}}$, we can simply replace the
limit 
\[
\lim_{\epsilon\to0}\left[X\left(z_{I}+2\epsilon\right)\bar{X}\left(\bar{z}_{I}+2\epsilon\right)\right]
\]
by $X\left(z_{I}\right)\bar{X}\left(\bar{z}_{I}\right)$. If $z_{I}=z_{I^{\left(r\right)}}$
and there are no $r^{\prime}\ne r$ such that $z_{I}=z_{I^{\left(r^{\prime}\right)}}$,
we obtain 
\begin{eqnarray}
 &  & \lim_{\epsilon\to0}\left[X\left(z_{I}+2\epsilon\right)\bar{X}\left(\bar{z}_{I}+2\epsilon\right)\right]\oint_{C_{r,\left|\epsilon\right|}}\frac{d\mathbf{z}}{2\pi i}\mathcal{S}\left(\mathbf{z},Z_{r}\right)\oint_{\bar{C}_{r,\left|\epsilon\right|}}\frac{d\bar{\mathbf{z}}}{2\pi i}\bar{\mathcal{S}}\left(\bar{\mathbf{z}},\bar{Z}_{r}\right)c\bar{c}e^{-\phi-\bar{\phi}}V_{r}^{\mathrm{DDF}}(Z_{r},\bar{Z}_{r})\nonumber \\
 &  & \quad=\left[\oint_{z_{I}}\frac{d\mathbf{z}}{2\pi i}\mathcal{S}\left(\mathbf{z},Z_{r}\right)X\left(z_{I}\right)-\frac{Q^{2}}{2}\frac{1}{p_{r}^{+}}e^{\phi}\partial\left(\psi^{+}e^{-i\frac{Q^{2}}{\alpha_{r}}X_{L}^{+}}\right)\left(z_{I}\right)\right]\nonumber \\
 &  & \hphantom{\quad=}\times\left[\oint_{\bar{z}_{I}}\frac{d\bar{\mathbf{z}}}{2\pi i}\bar{\mathcal{S}}\left(\bar{\mathbf{z}},\bar{Z}_{r}\right)\bar{X}\left(\bar{z}_{I}\right)-\frac{Q^{2}}{2}\frac{1}{p_{r}^{+}}e^{\bar{\phi}}\partial\left(\bar{\psi}^{+}e^{-i\frac{Q^{2}}{\alpha_{r}}X_{R}^{+}}\right)\left(\bar{z}_{I}\right)\right]\nonumber \\
 &  & \hphantom{\quad=}\times c\bar{c}e^{-\phi-\bar{\phi}}V_{r}^{\mathrm{DDF}}(Z_{r},\bar{Z}_{r})\,.\label{eq:XS1}
\end{eqnarray}
If $z_{I}=z_{I^{\left(r\right)}}=z_{I^{\left(r^{\prime}\right)}}$,
we get 
\begin{eqnarray}
 &  & \lim_{\epsilon\to0}\left[X\left(z_{I}+2\epsilon\right)\bar{X}\left(\bar{z}_{I}+2\epsilon\right)\right]\oint_{C_{r,\left|\epsilon\right|}}\frac{d\mathbf{z}}{2\pi i}\mathcal{S}\left(\mathbf{z},Z_{r}\right)\oint_{\bar{C}_{r,\left|\epsilon\right|}}\frac{d\bar{\mathbf{z}}}{2\pi i}\bar{\mathcal{S}}\left(\bar{\mathbf{z}},\bar{Z}_{r}\right)c\bar{c}e^{-\phi-\bar{\phi}}V_{r}^{\mathrm{DDF}}(Z_{r},\bar{Z}_{r})\nonumber \\
 &  & \hphantom{
X\left(z_{I}+2\epsilon\right)\bar{X}\left(\bar{z}_{I}+2\epsilon\right)%
}\times\oint_{C_{r^{\prime},\left|\epsilon\right|}}\frac{d\mathbf{z}}{2\pi i}\mathcal{S}\left(\mathbf{z},Z_{r^{\prime}}\right)\oint_{\bar{C}_{r^{\prime},\left|\epsilon\right|}}\frac{d\bar{\mathbf{z}}}{2\pi i}\bar{\mathcal{S}}\left(\bar{\mathbf{z}},\bar{Z}_{r^{\prime}}\right)c\bar{c}e^{-\phi-\bar{\phi}}V_{r^{\prime}}^{\mathrm{DDF}}(Z_{r^{\prime}},\bar{Z}_{r^{\prime}})\nonumber \\
 &  & \quad=\left[\oint_{z_{I}}\frac{d\mathbf{z}}{2\pi i}\mathcal{S}\left(\mathbf{z},Z_{r}\right)\oint_{z_{I}}\frac{d\mathbf{z}}{2\pi i}\mathcal{S}\left(\mathbf{z},Z_{r^{\prime}}\right)X\left(z_{I}\right)\vphantom{\left(\bar{\psi}^{+}e^{-i\frac{Q^{2}}{\alpha_{r}}X_{R}^{+}}\right)}\right.\nonumber \\
 &  & \hphantom{\quad=\qquad}-\frac{Q^{2}}{2}\frac{1}{p_{r}^{+}}\left\{ e^{\phi}\partial\left(\psi^{+}e^{-i\frac{Q^{2}}{\alpha_{r}}X_{L}^{+}}\right)\right\} \left(z_{I}\right)\oint_{z_{I}}\frac{d\mathbf{z}}{2\pi i}\mathcal{S}\left(\mathbf{z},Z_{r^{\prime}}\right)\nonumber \\
 &  & \hphantom{\quad=\qquad}\left.-\oint_{z_{I}}\frac{d\mathbf{z}}{2\pi i}\mathcal{S}\left(\mathbf{z},Z_{r}\right)\frac{Q^{2}}{2}\frac{1}{p_{r^{\prime}}^{+}}e^{\phi}\partial\left(\psi^{+}e^{-i\frac{Q^{2}}{\alpha_{r}}X_{L}^{+}}\right)\left(z_{I}\right)\right]\nonumber \\
 &  & \hphantom{\quad=}\times\left[\oint_{\bar{z}_{I}}\frac{d\bar{\mathbf{z}}}{2\pi i}\bar{\mathcal{S}}\left(\bar{\mathbf{z}},\bar{Z}_{r}\right)\oint_{\bar{z}_{I}}\frac{d\bar{\mathbf{z}}}{2\pi i}\bar{\mathcal{S}}\left(\bar{\mathbf{z}},\bar{Z}_{r^{\prime}}\right)\bar{X}\left(\bar{z}_{I}\right)\vphantom{\left(\bar{\psi}^{+}e^{-i\frac{Q^{2}}{\alpha_{r}}X_{R}^{+}}\right)}\right.\nonumber \\
 &  & \hphantom{\quad=\qquad\times}-\frac{Q^{2}}{2}\frac{1}{p_{r}^{+}}\left\{ e^{\bar{\phi}}\partial\left(\bar{\psi}^{+}e^{-i\frac{Q^{2}}{\alpha_{r}}X_{R}^{+}}\right)\right\} \left(\bar{z}_{I}\right)\oint_{\bar{z}_{I}}\frac{d\bar{\mathbf{z}}}{2\pi i}\bar{\mathcal{S}}\left(\bar{\mathbf{z}},\bar{Z}_{r^{\prime}}\right)\nonumber \\
 &  & \hphantom{\quad=\qquad\times}\left.-\oint_{\bar{z}_{I}}\frac{d\bar{\mathbf{z}}}{2\pi i}\bar{\mathcal{S}}\left(\bar{\mathbf{z}},\bar{Z}_{r}\right)\frac{Q^{2}}{2}\frac{1}{p_{r^{\prime}}^{+}}e^{\bar{\phi}}\partial\left(\bar{\psi}^{+}e^{-i\frac{Q^{2}}{\alpha_{r'}}X_{R}^{+}}\right)\left(\bar{z}_{I}\right)\right]\nonumber \\
 &  & \hphantom{\quad=}\times c\bar{c}e^{-\phi-\bar{\phi}}V_{r}^{\mathrm{DDF}}(Z_{r},\bar{Z}_{r})c\bar{c}e^{-\phi-\bar{\phi}}V_{r^{\prime}}^{\mathrm{DDF}}(Z_{r^{\prime}},\bar{Z}_{r^{\prime}})\,.\label{eq:XS2}
\end{eqnarray}
In the generic situation in which $z_{I},Z_{r}$ are all distinct,\footnote{The situation in which some of $z_{I},Z_{r}$ coincide can be considered
as a limit of these generic situations. } it is not possible for a $z_{I}$ to be equal to $z_{I^{\left(r\right)}}$
for more than two $r$'s, except for the tree-level three point amplitudes.
It is possible to derive the formula as (\ref{eq:XS2}) in such cases.
Thus we can see that the limit $\epsilon\to0$ in (\ref{eq:superBRSTinvariant2})
is not singular and we denote the result by the naive expression (\ref{eq:superBRSTinvariant}).
Eq.(\ref{eq:superBRSTinvariant}) is proved in appendix \ref{sec:A-proof-of}.

When $Q=0$, (\ref{eq:superBRSTinvariant}) becomes 
\begin{eqnarray}
A_{N}^{(g)} & \propto & \int[dT][d\alpha][\alpha d\theta]\nonumber \\
 &  & \times\int\left[dX^{\mu}d\psi^{\mu}d\bar{\psi}^{\mu}dbd\bar{b}dcd\bar{c}d\beta d\bar{\beta}d\gamma d\bar{\gamma}\right]_{g_{z\bar{z}}^{\mathrm{A}}}e^{-S^{\mathrm{tot}}}\nonumber \\
 &  & \qquad\times\prod_{K=1}^{6g-6+2N}\left[\oint_{C_{K}}\frac{dz}{\partial\rho}b_{zz}+\varepsilon_{K}\oint_{\bar{C}_{K}}\frac{d\bar{z}}{\bar{\partial}\bar{\rho}}b_{\bar{z}\bar{z}}\right]\prod_{I}\left[X\left(z_{I}\right)\bar{X}\left(\bar{z}_{I}\right)\right]\nonumber \\
 &  & \qquad\times\prod_{r=1}^{N}\left[c\bar{c}e^{-\phi-\bar{\phi}}V_{r}^{\mathrm{DDF}}(Z_{r},\bar{Z}_{r})\right]\,.\label{eq:superBRSTinvariant10}
\end{eqnarray}
This expression coincides with the one obtained from the first-quantized
formalism putting the PCO's at the interaction points of the light-cone
diagram, although this expression suffers from the contact term divergences. 

\section{The amplitudes from the first-quantized formalism\label{sec:The-amplitudes-from}}

In recent papers \cite{Sen2015a,Sen2015b,Sen2015}, a way to calculate
superstring amplitudes using the PCO's is established. We would like to
show that the amplitudes calculated from the light-cone gauge string
field theory using the dimensional regularization coincide with those
obtained by the method of these papers.

\subsection{The prescription }

In this subsection, we will just briefly explain the prescription
given in \cite{Sen2015}.

In the first-quantized approach using PCO's, an amplitude is expressed
by an integral of a correlation function with a fixed number $K$
of PCO insertions over the relevant moduli space $M$ which is assumed
to have real dimension $n$. Each point $\dot{m\in M}$ determines
a Riemann surface $\Sigma(m)$. Let $Y$ be a fiber bundle over $M$
whose fiber is 
\[
\underbrace{\Sigma(m)\times\Sigma(m)\times\cdots\times\Sigma(m)}_{K\mbox{ times}}\,,
\]
in order to describe the insertions of $K$ PCO's, and $X$ be the
subspace of $Y$ which is obtained by omitting the bad points where
the spurious singularities arise. A point in $X$ is denoted by $(m;a)$
with $m\in M,\,a\in\Sigma(m)\times\Sigma(m)\times\cdots\times\Sigma(m)$
and define a map $\varphi:X\to M$ which maps $(m;a)$ to $m$.

If there existed a global section $s$ of $\varphi:X\to M$, the amplitude
would be given by 
\[
\int_{M}\omega_{n}(m;s(m))\,,
\]
where the integrand is schematically expressed as \cite{Sen2015a}
\begin{equation}
\omega_{n}(m;z_{1},\cdots,z_{K})\propto\left\langle \prod_{i=1}^{K}\left(X(z_{i})-\partial\xi(z_{i})dz_{i}\right)\prod_{s=1}^{n}\left(\int d^{2}\sigma\frac{\partial(\sqrt{g}g_{ij})}{\partial m_{s}}b^{ij}dm_{s}\right)\prod_{r=1}^{N}V_{r}\right\rangle _{n}\,.
\end{equation}
$\left\langle \quad\right\rangle $ here denotes the correlation function
of the worldsheet theory on $\Sigma(m)$, $V_{1},\cdots,V_{N}$ are
BRST invariant vertex operators, $m_{1},\cdots,m_{n}$ are the
coordinates of $M$ and the subscript $n$ denotes that we should
extract the $n$-form part of this expression.

\begin{figure}
\begin{centering}
\includegraphics[scale=0.5]{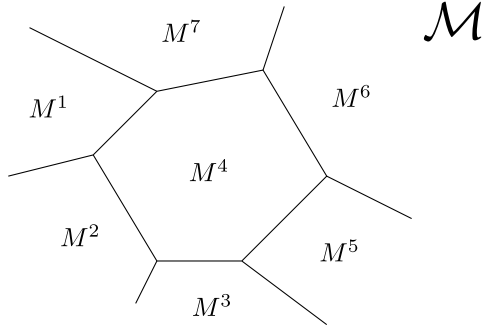} 
\par\end{centering}
\caption{Dual triangulation $\Upsilon$}
\end{figure}

Unfortunately such a global section does not exist in general. If
we divide $M$ into patches, we may be able to have a local section
on each patch. As was demonstrated in detail in \cite{Sen2015}, it
has been shown that 
\begin{enumerate}
\item One can pick a dual triangulation $\Upsilon$ of $M$ such that the
map $\varphi:X\to M$ has a local section $s^{\alpha}$ over each
of the codimension $0$ polyhedron $M^{\alpha}$ in $\Upsilon$. 
\item The amplitude can be given as 
\[
\sum_{\alpha}\int_{M^{\alpha}}\omega_{n}(m;s^{\alpha}(m))+A_{\mathrm{vertical}}\,,
\]
where $A_{\mathrm{vertical}}$ is the contribution of the ``vertical
integration''. $A_{\mathrm{vertical}}$ is given as a sum of integrals
of correlation functions involving $\xi,X$, antighost insertions
and the vertex operators, over $\partial M^{\alpha}$ and their submanifolds
which are called the vertical segments. 
\item The amplitude thus defined is independent of the choices of $\Upsilon,s^{\alpha}$
and the vertical segments, as long as the bad points are avoided. 
\item The amplitude thus defined is gauge invariant. 
\end{enumerate}

\subsection{$Q\to0$ limit of the light-cone gauge amplitudes }

As has been shown in \cite{Ishibashi2016a}, the Feynman amplitude
(\ref{eq:ANg}) is well-defined for $Q^{2}>10$ and we can define
$A_{N}^{(g)}$ as an analytic function of $Q$, i.e.\ $A_{N}^{(g)}(Q)$.
If the limit $Q\to0$ can be taken without encountering any divergences,
we will obtain the amplitude without the dilaton background. We would
like to show that the limit $Q\to0$ is smooth if there are no divergences
coming from the boundaries of the moduli space.

As has been shown in section \ref{sec:BRST-invariant-form}, the amplitude
(\ref{eq:ANg}) can be rewritten into a BRST invariant form (\ref{eq:superBRSTinvariant}).
Since the worldsheet theory used in (\ref{eq:superBRSTinvariant})
consists of the matter superconformal field theory with $\hat{c}=10$
and the superconformal ghosts, the amplitude (\ref{eq:superBRSTinvariant})
can be recast into the form described in the previous subsection as
\begin{equation}
A_{N}^{(g)}(Q)=\int_{M}\omega_{n}(m;s(m))\,,
\end{equation}
where 
\begin{eqnarray}
\omega_{n}(m;s(m)) & = & dm^{1}\wedge dm^{2}\wedge\cdots\wedge dm^{6g-6+2N}\nonumber \\
 &  & \times\int\left[dX^{\mu}d\psi^{\mu}d\bar{\psi}^{\mu}dbd\bar{b}dcd\bar{c}d\beta d\bar{\beta}d\gamma d\bar{\gamma}\right]_{g_{z\bar{z}}^{\mathrm{A}}}e^{-S^{\mathrm{tot}}}\nonumber \\
 &  & \ \ \ \times\prod_{K=1}^{6g-6+2N}\left[\oint_{C_{K}}\frac{dz}{\partial\rho}b_{zz}+\varepsilon_{K}\oint_{\bar{C}_{K}}\frac{d\bar{z}}{\bar{\partial}\bar{\rho}}b_{\bar{z}\bar{z}}\right]\prod_{I}\left[X\left(z_{I}\right)\bar{X}\left(\bar{z}_{I}\right)\right]\nonumber \\
 &  & \ \ \ \times\prod_{r=1}^{N}\left[\oint_{z_{I^{(r)}}}\frac{d\mathbf{z}}{2\pi i}\mathcal{S}\left(\mathbf{z},Z_{r}\right)\oint_{\bar{z}_{I^{(r)}}}\frac{d\bar{\mathbf{z}}}{2\pi i}\bar{\mathcal{S}}\left(\bar{\mathbf{z}},\bar{Z}_{r}\right)c\bar{c}e^{-\phi-\bar{\phi}}V_{r}^{\mathrm{DDF}}(Z_{r},\bar{Z}_{r})\right].~~~~~~
\end{eqnarray}
The section $s(m)$ here corresponds to the prescription where the
PCO's are located at the interaction points of the light-cone diagram.
This expression is well-defined for $Q^{2}>10$ but may suffer from
the spurious singularities otherwise.

It is also possible to define amplitudes with the same worldsheet
theory and the vertex operators, avoiding the spurious singularities
by the Sen-Witten prescription. Namely we can define 
\begin{equation}
A_{N}^{(g)\mathrm{SW}}(Q)=\sum_{\alpha}\int_{M^{\alpha}}\omega_{n}(m;s^{\prime\alpha}(m))+A_{\mathrm{vertical}}\,,
\end{equation}
with the local sections $s^{\prime\alpha}(m)$, avoiding the spurious
singularities for $Q^{2}<10$. For $Q^{2}>10$, 
\begin{equation}
A_{N}^{(g)}(Q)=A_{N}^{(g)\mathrm{SW}}(Q)\,,
\end{equation}
because there are no bad points for $Q^{2}>10$ and the results do
not depend on the choice of the local sections. Therefore as an analytic
function of $Q$, $A_{N}^{(g)}(Q)$ coincides with $A_{N}^{(g)\mathrm{SW}}(Q)$
and we obtain 
\begin{equation}
\lim_{Q\to0}A_{N}^{(g)}(Q)=A_{N}^{(g)\mathrm{SW}}(0)\,,
\end{equation}
if the right hand side exists. The Sen-Witten prescription deals with
the spurious singularities and if the superstring amplitudes in question
does not suffer from the infrared divergences, $A_{N}^{(g)\mathrm{SW}}(0)$
is well-defined. We have shown that the Feynman amplitudes calculated
by our method coincide with those obtained by the first-quantized
method using the Sen-Witten prescription, as long as we consider infrared
safe quantities.

\section{Conclusions and discussions\label{sec:Conclusions-and-discussions}}

In this paper, we have studied the regularization of the contact term
divergences of multiloop scattering amplitudes in the light-cone gauge
superstring field theory. We have used the theory in a linear dilaton
background $\Phi=-iQX^{1}$. The divergences of the amplitudes of
the theory are thoroughly analyzed in \cite{Ishibashi2016a}. Since
the central charge of this theory is $c=12-12Q^{2}$, it is possible
to shift the central charge to a large negative value by putting $Q^{2}$
large. The multiloop amplitudes involving only the even spin structure
with the external lines in the NS-NS sector are indeed regularized
by taking $Q^{2}>10.$ We have shown that the resultant amplitudes
coincide with those of the first-quantized theory through the analytic
continuation $Q\to0$, without encountering the divergences except
those originating from the boundaries of the moduli space. Similarly
to the dimensional regularization previously considered \cite{Baba:2009ns,Baba:2009fi,Baba:2009zm,Ishibashi:2010nq},
the amplitudes can be recast into a BRST invariant form of the worldsheet
theory in the conformal gauge. This can be achieved by adding the
supersymmetric $X^{\pm}$ CFT for the longitudinal variables and the
superreparametrization ghosts to the worldsheet theory. In the present
case, we have constructed BRST invariant worldsheet theory by setting
${\displaystyle \frac{d-10}{8}=-Q^{2}}$ in the action of the $X^{\pm}$
CFT given in (\ref{eq:superSXpm}).

In order to make our regularization scheme complete, we need to deal
with the amplitudes for odd spin structure and those with external
lines in the Ramond sector. Contrary to the dimensional regularization
in which the number of the transverse dimensions $d-2$ is naively
shifted, it remains $8$ in the present case. This implies that the
present procedure does not give rise to the problem pointed out in
\cite{Ishibashi:2010nq} in constructing the space-time fermions.
Furthermore, the light-cone gauge theory in the linear dilaton background
used in this paper is much simpler than the theory proposed in \cite{Ishibashi:2011fy}.
We will investigate this extension elsewhere.

With our prescription, we may be able to describe superstring theory
by the simple action with only three string vertices. However there
exists subtle points in such a formulation. The action of the light-cone
gauge closed superstring field theory possesses only the cubic interactions.
While this fact makes the theory simple, the Hamiltonian is unbounded
below and thus unstable. The light-cone gauge theory does not contain
auxiliary fields and hence the cubic interactions are considered to
directly mean the instability of the perturbative vacuum. However,
we have to pay attention to the fact that the point where $p^{+}=0$
is not regular in the light-cone gauge formulation and the nonperturbative
properties reside there. The stability of the vacuum might not be
such a simple problem in the light-cone gauge closed superstring field
theory. These facts suggest that it would be desirable to have a gauge
invariant string field theory to which our method here is applicable.
The conformal gauge expression of the amplitudes given in section
\ref{sec:BRST-invariant-form} may give us a hint about how to construct
such a theory. We hope that we will also study these issues elsewhere.

\section*{Acknowledgments}

N.I. would like to thank Ted Erler and Ashoke Sen for useful comments.
He also would like to thank the organizers of ``VIII Workshop on
String Field Theory and Related Aspects'' at S\~ao Paulo, especially
N. Berkovits, for hospitality. K.M. would like to thank the hospitality
of Okayama Institute for Quantum Physics, where part of this work
was done. This work was supported in part by Grant-in-Aid for Scientific
Research (C) (25400242) and (15K05063) from MEXT.

\appendix

\section{Arakelov metric and Arakelov Green's function\label{sec:Arakelov-metric-and}}

The Arakelov metric $\hat{g}_{z\bar{z}}^{\mathrm{A}}$ and the Arakelov
Green's function $G^{\mathrm{A}}\left(z;w\right)$ are defined as
follows. Let $\mu_{z\bar{z}}$ be 
\begin{equation}
\mu_{z\bar{z}}\equiv\frac{1}{2g}\omega(z)\frac{1}{\mathrm{Im}\Omega}\bar{\omega}(\bar{z})~.\label{eq:Bergman}
\end{equation}
We note that 
\begin{equation}
\int_{\Sigma}dz\wedge d\bar{z}\,i\mu_{z\bar{z}}=1~,\label{eq:norm-Bergman}
\end{equation}
which follows from 
\begin{equation}
\int_{\Sigma}\omega_{\mu}\wedge\bar{\omega}_{\nu}=-2i\mathrm{Im}\Omega_{\mu\nu}~.
\end{equation}
The Arakelov metric on $\Sigma$, 
\begin{equation}
ds_{\mathrm{A}}^{\;2}=2g_{z\bar{z}}^{\mathrm{A}}dzd\bar{z}~,\label{eq:Arakelov}
\end{equation}
is defined so that its scalar curvature $R^{\mathrm{A}}\equiv-2g^{\mathrm{A}z\bar{z}}\partial\bar{\partial}\ln g_{z\bar{z}}^{\mathrm{A}}$
satisfies 
\begin{equation}
g_{z\bar{z}}^{\mathrm{A}}R^{\mathrm{A}}=-8\pi(g-1)\mu_{z\bar{z}}~.\label{eq:ArakelovR}
\end{equation}
This condition determines $g_{z\bar{z}}^{\mathrm{A}}$ only up to
an overall constant, which will be chosen later.

The Arakelov Green's function $G^{\mathrm{A}}(z,\bar{z};w,\bar{w})$
with respect to the Arakelov metric is defined to satisfy\footnote{The delta function $\delta^{2}(z-w)$ is normalized by ${\displaystyle \int dz\wedge d\bar{z}}\,i\delta^{2}(z-w)=1$.}
\begin{eqnarray}
 &  & -\partial_{z}\partial_{\bar{z}}G^{\mathrm{A}}(z,\bar{z};w,\bar{w})=2\pi\delta^{2}(z-w)-2\pi\mu_{z\bar{z}}~,\nonumber \\
 &  & \int_{\Sigma}dz\wedge d\bar{z}\,i\mu_{z\bar{z}}G^{\mathrm{A}}(z,\bar{z};w,\bar{w})=0~.\label{eq:G-B}
\end{eqnarray}
One can obtain a more explicit form of $G^{\mathrm{A}}(z,\bar{z};w,\bar{w})$
by solving (\ref{eq:G-B}) for $G^{\mathrm{A}}(z,\bar{z};w,\bar{w})$.
Let $F(z,\bar{z};w,\bar{w})$ be the $\left(-\frac{1}{2},-\frac{1}{2}\right)\times\left(-\frac{1}{2},-\frac{1}{2}\right)$
form on $\Sigma\times\Sigma$ which satisfies 
\begin{equation}
\partial_{z}\partial_{\bar{z}}\ln F(z,\bar{z};w,\bar{w})=2\pi\delta^{2}(z-w)-2\pi g\mu_{z\bar{z}}~,\label{eq:diffeq-propagator}
\end{equation}
which can be given by 
\begin{equation}
F(z,\bar{z};w,\bar{w})=\exp\left[-2\pi\mathrm{Im}\int_{w}^{z}\omega\frac{1}{\mathrm{Im}\Omega}\mathrm{Im}\int_{w}^{z}\omega\right]\left|E(z,w)\right|^{2}~.\label{eq:propagator2}
\end{equation}
Putting (\ref{eq:diffeq-propagator}) and (\ref{eq:ArakelovR}) together,
we find that $G^{\mathrm{A}}(z,\bar{z};w,\bar{w})$ is given by 
\begin{equation}
G^{\mathrm{A}}(z,\bar{z};w,\bar{w})=-\ln F(z,\bar{z};w,\bar{w})-\frac{1}{2}\ln\left(2g_{z\bar{z}}^{\mathrm{A}}\right)-\frac{1}{2}\ln\left(2g_{w\bar{w}}^{\mathrm{A}}\right)~,\label{eq:G-F-g-g}
\end{equation}
up to an additive constant independent of $z,\bar{z}$ and $w,\bar{w}$.
This possible additive constant can be absorbed into the ambiguity
in the overall constant of $g_{z\bar{z}}^{\mathrm{A}}$ mentioned
above. It is required that $\eqref{eq:G-F-g-g}$ holds exactly as
it is \cite{Dugan:1987qe,Sonoda:1987ra,D'Hoker:1989ae}. This implies
that 
\begin{equation}
2g_{z\bar{z}}^{\mathrm{A}}=\lim_{w\to z}\exp\left[-G^{\mathrm{A}}(z,\bar{z};w,\bar{w})-\ln|z-w|^{2}\right]~,\label{eq:gA-expGA}
\end{equation}
and the overall constant of $g_{z\bar{z}}^{\mathrm{A}}$ is, in principle,
determined by the second relation in (\ref{eq:G-B}).

\section{Supersymmetric $X^{\pm}$ CFT\label{sec:Supersymmetric--CFT}}

The conformal gauge worldsheet theory corresponding to the light-cone
gauge superstring theory in noncritical dimensions was studied in
\cite{Baba:2009fi}. The longitudinal part of it is called the supersymmetric
$X^{\pm}$ CFT whose action is given by 
\begin{equation}
S_{\mathrm{super}}^{\pm}\left[\mathcal{X}^{\pm};\hat{g}_{z\bar{z}}\right]=-\frac{1}{2\pi}\int d^{2}\mathbf{z}\left(\bar{D}\mathcal{X}^{+}D\mathcal{X}^{-}+\bar{D}\mathcal{X}^{-}D\mathcal{X}^{+}\right)+\frac{d-10}{8}\Gamma_{\mathrm{super}}\left[\mathcal{X}^{+};\hat{g}_{z\bar{z}}\right]\,.\label{eq:superSXpm}
\end{equation}
Here the supercoordinate $\mathbf{z}$ is given by 
\begin{equation}
\mathbf{z}=(z,\theta)\,,
\end{equation}
the superfield $\mathcal{X}^{\pm}$ is defined as 
\begin{equation}
\mathcal{X}^{\pm}\left(\mathbf{z},\bar{\mathbf{z}}\right)=X^{\pm}\left(z\right)+i\theta\psi^{\pm}\left(z\right)+i\bar{\theta}\bar{\psi}^{\pm}\left(\bar{z}\right)+\theta\bar{\theta}F^{\pm}\,,\label{eq:superfieldXpm}
\end{equation}
and 
\begin{eqnarray}
D & \equiv & \frac{\partial}{\partial\theta}+\theta\frac{\partial}{\partial z}\,,\nonumber \\
\bar{D} & \equiv & \frac{\partial}{\partial\bar{\theta}}+\bar{\theta}\frac{\partial}{\partial\bar{z}}\,,\nonumber \\
d^{2}\mathbf{z} & \equiv & d\left(\mathrm{Re}z\right)d\left(\mathrm{Im}z\right)d\theta d\bar{\theta}\,.
\end{eqnarray}
The interaction term $\Gamma_{\mathrm{super}}$ is given by 
\begin{eqnarray}
\Gamma_{\mathrm{super}}\left[\mathcal{X}^{+};\hat{g}_{z\bar{z}}\right] & = & -\frac{1}{2\pi}\int d^{2}\mathbf{z}\left(\bar{D}\Phi D\Phi+\theta\bar{\theta}\hat{g}_{z\bar{z}}\hat{R}\Phi\right)\,,\nonumber \\
\Phi\left(\mathbf{z},\bar{\mathbf{z}}\right) & = & \ln\left(\left(D\Theta^{+}\right)^{2}\left(\mathbf{z}\right)\left(\bar{D}\bar{\Theta}^{+}\right)^{2}\left(\bar{\mathbf{z}}\right)\right)-\ln\hat{g}_{z\bar{z}}\,,\label{eq:Phi}\\
\Theta^{+}\left(\mathbf{z}\right) & = & \frac{D\mathcal{X}^{+}}{(\partial\mathcal{X}^{+})^{\frac{1}{2}}}\left(\mathbf{z}\right)\,,\nonumber 
\end{eqnarray}
which is the super Liouville action defined for variable $\Phi$ with
the background metric $ds^{2}=2\hat{g}_{z\bar{z}}dzd\bar{z}$. The
super energy-momentum tensor $T^{\mathcal{X}^{\pm}}(\mathbf{z})$
becomes 
\begin{equation}
T^{\mathcal{X}^{\pm}}(\mathbf{z})=\frac{1}{2}\partial\mathcal{X}^{+}D\mathcal{X}^{-}+\frac{1}{2}\partial\mathcal{X}^{-}D\mathcal{X}^{+}-\frac{d-10}{4}S(\mathbf{z},\mbox{\mathversion{bold}\ensuremath{\mathcal{X}_{L}^{+}}})\,,
\end{equation}
where $S(\mathbf{z},\mbox{\mathversion{bold}\ensuremath{\mathcal{X}_{L}^{+}}})$
denotes the super Schwarzian derivative 
\begin{equation}
S(\mathbf{z},\mbox{\mathversion{bold}\ensuremath{\mathcal{X}_{L}^{+}}})=\frac{\partial^{2}\Theta^{+}}{D\Theta}-2\frac{\partial D\Theta^{+}\partial\Theta^{+}}{\left(D\Theta^{+}\right)^{2}}~.
\end{equation}

In the present context, for transverse sector we use the superstring
theory in a linear dilaton background instead of the theory in noncritical
dimensions. We therefore consider the theory (\ref{eq:superSXpm})
with the identification ${\displaystyle \frac{d-10}{8}=-Q^{2}}.$

The correlation functions to be considered in this theory are defined
as 
\begin{eqnarray}
 &  & \left\langle \prod_{r=1}^{N}e^{-ip_{r}^{+}\mathcal{X}^{-}}(\mathbf{Z}_{r},\bar{\mathbf{Z}}_{r})\prod_{s=1}^{M}e^{-ip_{s}^{-}\mathcal{X}^{+}}(\mathbf{w}_{s},\bar{\mathbf{w}}_{s})\right\rangle _{\hat{g}_{z\bar{z}}}^{\mathcal{X}^{\pm}}\nonumber \\
 &  & \quad\equiv Z_{\mathrm{super}}^{\mathcal{X}}[\hat{g}_{z\bar{z}}]^{-2}\int\left[d\mathcal{X}^{+}d\mathcal{X}^{-}\right]_{\hat{g}_{z\bar{z}}}e^{-S_{\mathrm{super}}^{\pm}\left[\mathcal{X}^{\pm};\hat{g}_{z\bar{z}}\right]}\prod_{r=1}^{N}e^{-ip_{r}^{+}\mathcal{X}^{-}}(\mathbf{Z}_{r},\bar{\mathbf{Z}}_{r})\prod_{s=1}^{M}e^{-ip_{s}^{-}\mathcal{X}^{+}}(\mathbf{w}_{s},\bar{\mathbf{w}}_{s})\,.\nonumber \\
\label{eq:superXpmcorr}
\end{eqnarray}
Here 
\begin{equation}
Z_{\mathrm{super}}^{\mathcal{X}}[\hat{g}_{z\bar{z}}]=\int\left[d\mathcal{X}\right]_{\hat{g}_{z\bar{z}}}\exp\left[-\frac{1}{2\pi}\int d^{2}\mathbf{z}\bar{D}\mathcal{X}D\mathcal{X}\right]\,,
\end{equation}
and 
\begin{eqnarray}
\mathbf{Z}_{r} & = & (Z_{r},\Theta_{r})\,,\nonumber \\
\mathbf{w}_{s} & = & (w_{s},\eta_{s})\,.
\end{eqnarray}
In order to discuss these correlation functions, it is convenient
to introduce supersymmetric version of $\rho(z)$ in (\ref{eq:mandelstammulti})
which is defined by 
\begin{equation}
\rho_{s}\left(\mathbf{z}\right)=\rho\left(z\right)+\theta f(z)\,,
\end{equation}
where 
\begin{equation}
f(z)\equiv-\sum_{r}\alpha_{r}\Theta_{r}S_{\delta}\left(z,Z_{r}\right)\,,
\end{equation}
and $S_{\delta}\left(z,z^{\prime}\right)$ is the fermion's Green's
function corresponding to the spin structure $\delta$. When all the
external lines are in the NS-NS sector and $\delta$ is an even spin
structure, $S_{\delta}\left(z,w\right)$ is equal to the so-called
Szego kernel 
\begin{equation}
S_{\delta}\left(z,w\right)=\frac{1}{E\left(z,w\right)}\frac{\vartheta\left[\delta\right]\left(\left.\int_{w}^{z}\omega\right|\Omega\right)}{\vartheta\left[\delta\right]\left(0|\Omega\right)}\,,
\end{equation}
where $\vartheta[\delta](\zeta|\Omega)$ denotes the theta function
with characteristics $[\delta]={\delta^{\prime}\atopwithdelims[]\delta^{\prime\prime}}$,
namely $\delta=\Omega\delta'+\delta''$ $\left(\delta',\delta''\in\mathbb{R}^{g}\right)$,
given by 
\begin{equation}
\vartheta[\delta](\zeta|\Omega)=\sum_{n\in\mathbb{Z}^{g}}e^{2\pi i\left[\frac{1}{2}(n+\delta^{\prime})\Omega(n+\delta^{\prime})+(n+\delta^{\prime})(\zeta+\delta^{\prime\prime})\right]}\,,
\end{equation}
for $\zeta\in\mathbb{C}^{g}/(\mathbb{Z}^{g}+\Omega\mathbb{Z}^{g})$
. The right hand side of (\ref{eq:superXpmcorr}) can be calculated
to be 
\begin{eqnarray}
 &  & \left\langle \prod_{r=1}^{N}e^{-ip_{r}^{+}\mathcal{X}^{-}}(\mathbf{Z}_{r},\bar{\mathbf{Z}}_{r})\prod_{s=1}^{M}e^{-ip_{s}^{-}\mathcal{X}^{+}}(\mathbf{w}_{s},\bar{\mathbf{w}}_{s})\right\rangle _{\hat{g}_{z\bar{z}}}^{\mathcal{X}^{\pm}}\nonumber \\
 &  & \quad=(2\pi)^{2}\delta\left(\sum_{s}p_{s}^{-}\right)\delta\left(\sum_{r}p_{r}^{+}\right)\prod_{s}e^{-p_{s}^{-}\frac{\mathbf{\rho}_{s}+\bar{\mathbf{\rho}}_{s}}{2}}(\mathbf{w}_{s},\bar{\mathbf{w}}_{s})\,e^{-\frac{d-10}{8}\Gamma_{\mathrm{super}}\left[\rho_{s}+\bar{\rho}_{s};\hat{g}_{z\bar{z}}\right]}\,,~~~~~\label{eq:superXpmcorr2}
\end{eqnarray}
where 
\begin{equation}
\exp\left(-\Gamma_{\mathrm{super}}\left[\rho_{s}+\bar{\rho}_{s};\hat{g}_{z\bar{z}}\right]\right)=\exp\left(-\frac{1}{2}\Gamma\left[\sigma;\hat{g}_{z\bar{z}}\right]-\sum_{r}\Delta\Gamma_{r}-\sum_{I}\Delta\Gamma_{I}\right)\,,\label{eq:Gammasupermulti2}
\end{equation}
with
\begin{eqnarray}
-\Delta\Gamma_{r} & = & \frac{1}{2\alpha_{r}}\frac{\partial ff}{\partial^{2}\rho}\left(z_{I^{\left(r\right)}}\right)+\mathrm{c.c.}\,,\nonumber \\
-\Delta\Gamma_{I} & = & \left\{ -\left(\frac{5}{12}\frac{\partial^{4}\rho}{\left(\partial^{2}\rho\right)^{3}}-\frac{3}{4}\frac{\left(\partial^{3}\rho\right)^{2}}{\left(\partial^{2}\rho\right)^{4}}\right)\partial ff+\frac{2}{3}\frac{\partial^{3}ff}{\left(\partial^{2}\rho\right)^{2}}-\frac{\partial^{3}\rho}{\left(\partial^{2}\rho\right)^{3}}\partial^{2}ff\right.\nonumber \\
 &  & \left.\quad{}-\frac{1}{12}\frac{\partial^{3}f\partial^{2}f\partial ff}{\left(\partial^{2}\rho\right)^{4}}\right\} \left(z_{I}\right)+\mathrm{c.c.}\,,
\end{eqnarray}
and $\Gamma\left[\sigma;\hat{g}_{z\bar{z}}\right]$ given in (\ref{eq:e^-Gamma}).
From (\ref{eq:superXpmcorr2}), it is possible to deduce all the correlation
functions of the $X^{\pm}$ CFT.

\subsubsection*{The correlation functions of fermions}

From (\ref{eq:superXpmcorr2}), one can derive a formula of the correlation
functions of fermions $\psi^{\pm}, \bar{\psi}^{\pm}$, which is useful 
in appendix \ref{sec:A-proof-of}.
(\ref{eq:superXpmcorr2}) can be rewritten as\footnote{From the correlation functions, we can see that $F^{+}$ and $F^{-}$
can be put equal to zero. } 
\begin{eqnarray}
 &  & \left\langle \prod_{r=1}^{N}e^{-ip_{r}^{+}X^{-}+p_{r}^{+}\left(\Theta_{r}\psi^{-}+\bar{\Theta}_{r}\bar{\psi}^{-}\right)}(Z_{r},\bar{Z}_{r})\prod_{s=1}^{M}e^{-ip_{s}^{-}X^{+}+p_{s}^{-}\left(\eta_{s}\psi^{+}+\bar{\eta}_{s}\bar{\psi}^{+}\right)}(w_{s},\bar{w}_{s})\right\rangle _{\hat{g}_{z\bar{z}}}^{\mathcal{X}^{\pm}}\nonumber \\
 &  & \quad=(2\pi)^{2}\delta\left(\sum_{s}p_{s}^{-}\right)\delta\left(\sum_{r}p_{r}^{+}\right)\prod_{s}e^{-p_{s}^{-}\frac{1}{2}\left(\mathbf{\rho}+\bar{\mathbf{\rho}}\right)}(w_{s},\bar{w}_{s})e^{-\frac{d-10}{16}\Gamma}\nonumber \\
 &  & \hphantom{\quad=\quad}\times\prod_{s}e^{-p_{s}^{-}\frac{1}{2}\left(\eta_{s}f\left(w_{s}\right)+\bar{\eta}_{s}\bar{f}\left(\bar{w}_{s}\right)\right)}e^{-\frac{d-10}{8}\left(\sum_{r}\Delta\Gamma_{r}+\sum_{I}\Delta\Gamma_{I}\right)}\,.\label{eq:superXpmcorr3}
\end{eqnarray}
The fermionic contribution of this correlation function can be expressed
as 
\begin{eqnarray}
 &  & \prod_{s}e^{-p_{s}^{-}\frac{1}{2}\left(\eta_{s}f\left(w_{s}\right)+\bar{\eta}_{s}\bar{f}\left(\bar{w}_{s}\right)\right)}e^{-\frac{d-10}{8}\left(\sum_{r}\Delta\Gamma_{r}+\sum_{I}\Delta\Gamma_{I}\right)}\nonumber \\
 &  & \quad=\left(Z^{\psi}[\hat{g}_{z\bar{z}}]\right)^{-2}\int\left[d\psi^{+}d\psi^{-}d\bar{\psi}^{+}d\bar{\psi}^{-}\right]e^{\frac{1}{\pi}\int d^{2}z\left(\psi^{-}\bar{\partial}\psi^{+}+\bar{\psi}^{-}\partial\bar{\psi}^{+}\right)-S_{\mathrm{int}}}\nonumber \\
 &  & \hphantom{\quad=\left(Z^{\psi}[\hat{g}_{z\bar{z}}]\right)^{-2}\quad}\times\prod_{r=1}^{N}e^{p_{r}^{+}\left(\Theta_{r}\psi^{-}+\bar{\Theta}_{r}\bar{\psi}^{-}\right)}(Z_{r},\bar{Z}_{r})\prod_{s=1}^{M}e^{p_{s}^{-}\left(\eta_{s}\psi^{+}+\bar{\eta}_{s}\bar{\psi}^{+}\right)}(w_{s},\bar{w}_{s})\,,\label{eq:fermioniccorr}
\end{eqnarray}
where 
\begin{eqnarray}
S_{\mathrm{int}} & = & \frac{d-10}{8}\left[-\sum_{r}\frac{2}{\alpha_{r}}\frac{\partial\psi^{+}\psi^{+}}{\partial^{2}\rho}\left(z_{I^{\left(r\right)}}\right)\vphantom{\left\{ \left(\frac{5}{3}\frac{\partial^{4}\rho}{\left(\partial^{2}\rho\right)^{3}}-3\frac{\left(\partial^{3}\rho\right)^{2}}{\left(\partial^{2}\rho\right)^{4}}\right)\partial\psi^{+}\psi^{+}-\frac{8}{3}\frac{\partial^{3}\psi^{+}\psi^{+}}{\left(\partial^{2}\rho\right)^{2}}\right.}\right.\nonumber \\
 &  & \hphantom{\frac{d-10}{8}\qquad}+\sum_{I}\left\{ \left(\frac{5}{3}\frac{\partial^{4}\rho}{\left(\partial^{2}\rho\right)^{3}}-3\frac{\left(\partial^{3}\rho\right)^{2}}{\left(\partial^{2}\rho\right)^{4}}\right)\partial\psi^{+}\psi^{+}-\frac{8}{3}\frac{\partial^{3}\psi^{+}\psi^{+}}{\left(\partial^{2}\rho\right)^{2}}\right.\nonumber \\
 &  & \hphantom{\frac{d-10}{8}\qquad+\sum_{I}\left(\frac{5}{3}\right)}\left.+\frac{4\partial^{3}\rho}{\left(\partial^{2}\rho\right)^{3}}\partial^{2}\psi^{+}\psi^{+}+\frac{4}{3}\frac{\partial^{3}\psi^{+}\partial^{2}\psi^{+}\partial\psi^{+}\psi^{+}}{\left(\partial^{2}\rho\right)^{4}}\right\} \left(z_{I}\right)\nonumber \\
 &  & \hphantom{\frac{d-10}{8}\quad}\left.+\mathrm{c.c.}\vphantom{\left\{ \left(\frac{5}{3}\frac{\partial^{4}\rho}{\left(\partial^{2}\rho\right)^{3}}-3\frac{\left(\partial^{3}\rho\right)^{2}}{\left(\partial^{2}\rho\right)^{4}}\right)\partial\psi^{+}\psi^{+}-\frac{8}{3}\frac{\partial^{3}\psi^{+}\psi^{+}}{\left(\partial^{2}\rho\right)^{2}}\right.}\right]\,.\label{eq:Sint}
\end{eqnarray}
Substituting (\ref{eq:fermioniccorr}) into (\ref{eq:superXpmcorr3})
and differentiating with respect to $\Theta_{r},\bar{\Theta}_{r},\eta_{s},\bar{\eta}_{s}$,
we obtain the following identity 
\begin{eqnarray}
 &  & \left\langle \prod_{r=1}^{N}e^{-ip_{r}^{+}X^{-}}(Z_{r},\bar{Z}_{r})\prod_{s=1}^{M}e^{-ip_{s}^{-}X^{+}}(w_{s},\bar{w}_{s})\psi^{+}\left(u_{1}\right)\cdots\psi^{+}\left(u_{n}\right)\psi^{-}\left(v_{1}\right)\cdots\psi^{-}\left(v_{m}\right)\right.\nonumber \\
 &  & \hphantom{\prod_{r=1}^{N}e^{-ip_{r}^{+}X^{-}}(Z_{r},\bar{Z}_{r})\prod_{s=1}^{M}e^{-ip_{s}^{-}X^{+}}(w_{s},\bar{w}_{s})}\left.\times\bar{\psi}^{+}\left(\tilde{u}_{1}\right)\cdots\bar{\psi}^{+}\left(\tilde{u}_{n}\right)\bar{\psi}^{-}\left(\tilde{v}_{1}\right)\cdots\bar{\psi}^{-}\left(\tilde{v}_{m}\right)\vphantom{\prod_{r=1}^{N}e^{-ip_{r}^{+}X^{-}}(Z_{r},\bar{Z}_{r})\prod_{s=1}^{M}e^{-ip_{s}^{-}X^{+}}(w_{s},\bar{w}_{s})}\right\rangle _{\hat{g}_{z\bar{z}}}^{\mathcal{X}^{\pm}}\nonumber \\
 &  & \quad=(2\pi)^{2}\delta\left(\sum_{s}p_{s}^{-}\right)\delta\left(\sum_{r}p_{r}^{+}\right)\prod_{s}e^{-p_{s}^{-}\frac{1}{2}\left(\mathbf{\rho}+\bar{\mathbf{\rho}}\right)}(w_{s},\bar{w}_{s})e^{-\frac{d-10}{16}\Gamma}\nonumber \\
 &  & \hphantom{\quad=\quad}\times\left(Z^{\psi}[\hat{g}_{z\bar{z}}]\right)^{-2}\int\left[d\psi^{+}d\psi^{-}d\bar{\psi}^{+}d\bar{\psi}^{-}\right]e^{\frac{1}{\pi}\int d^{2}z\left(\psi^{-}\bar{\partial}\psi^{+}+\bar{\psi}^{-}\partial\bar{\psi}^{+}\right)-S_{\mathrm{int}}}\nonumber \\
 &  & \hphantom{\left(Z^{\psi}[\hat{g}_{z\bar{z}}]\right)^{-2}\int\left[d\psi^{+}d\psi^{-}d\bar{\psi}^{+}d\bar{\psi}^{-}\right]\quad}\times\psi^{+}\left(u_{1}\right)\cdots\psi^{+}\left(u_{n}\right)\psi^{-}\left(v_{1}\right)\cdots\psi^{-}\left(v_{m}\right)\nonumber \\
 &  & \hphantom{\left(Z^{\psi}[\hat{g}_{z\bar{z}}]\right)^{-2}\int\left[d\psi^{+}d\psi^{-}d\bar{\psi}^{+}d\bar{\psi}^{-}\right]\quad}\times\bar{\psi}^{+}\left(\tilde{u}_{1}\right)\cdots\bar{\psi}^{+}\left(\tilde{u}_{n}\right)\bar{\psi}^{-}\left(\tilde{v}_{1}\right)\cdots\bar{\psi}^{-}\left(\tilde{v}_{m}\right)\,.\label{eq:fermioniccorr2}
\end{eqnarray}
Namely, the fermionic part of the correlation functions of supersymmetric
$X^{\pm}$ CFT coincide with those of the theory with interaction
$S_{\mathrm{int}}$ which is localized at the interaction points $z_{I}$.

\section{Ghost systems on higher genus Riemann surfaces\label{sec:-systems-on-higher}}

In this appendix, we would like to show some identities which are
crucial in deriving the conformal gauge expression of the light-cone
gauge amplitudes in section \ref{sec:BRST-invariant-form}.

Let us consider the conformal field theory with the action 
\begin{equation}
\frac{1}{\pi}\int dz\wedge d\bar{z}\sqrt{g}\left(b\nabla^{z}c+\bar{b}\nabla^{\bar{z}}\bar{c}\right)\,,
\end{equation}
where the fields $b,c$ are with conformal weight $\left(\lambda,0\right),\left(1-\lambda,0\right)$
and $\bar{b},\bar{c}$ are their antiholomorphic counterparts with
conformal weight $(0,\lambda),(0,1-\lambda)$. Here we consider the
case either $\lambda\in\mathbb{Z}$ or $\lambda\in\mathbb{Z}+\frac{1}{2}$.
The fields can be either Grassmann odd or even accordingly. We define
$\epsilon=\pm1$ to be 
\begin{equation}
\epsilon=\begin{cases}
1 & \mbox{if }b,c\mbox{ are Grassmann odd}\\
-1 & \mbox{if }b,c\mbox{ are Grassmann even}
\end{cases}\,.
\end{equation}
There exist local operators $e^{q\phi}\left(z,\bar{z}\right)\,\left(q\in\frac{\mathbb{Z}}{2}\right)$,
which satisfy 
\begin{eqnarray}
b\left(z\right)e^{q\phi}\left(w,\bar{w}\right) & \sim & \left(z-w\right)^{-\epsilon q}\,,\nonumber \\
c\left(z\right)e^{q\phi}\left(w,\bar{w}\right) & \sim & \left(z-w\right)^{\epsilon q}\,,\nonumber \\
\bar{b}\left(\bar{z}\right)e^{q\phi}\left(w,\bar{w}\right) & \sim & \left(\bar{z}-\bar{w}\right)^{-\epsilon q}\,,\nonumber \\
\bar{c}\left(\bar{z}\right)e^{q\phi}\left(w,\bar{w}\right) & \sim & \left(\bar{z}-\bar{w}\right)^{\epsilon q}\,.\label{eq:expqphi}
\end{eqnarray}

We would like to discuss the correlation functions of the form 
\begin{equation}
\left\langle \prod_{i}e^{\epsilon q_{i}\phi}\left(z_{i},\bar{z}_{i}\right)\right\rangle \,,
\end{equation}
on a genus $g$ Riemann surface. $q_{i}$ should satisfy 
\begin{equation}
\sum_{i}q_{i}=-\left(2\lambda-1\right)\left(g-1\right)\,.
\end{equation}
When $\lambda\in\mathbb{Z}+\frac{1}{2}$, the correlation function
we consider here is the one corresponding to a spin structure $\alpha^{\prime}\atopwithdelims[]\alpha^{\prime\prime}$.
Namely, the fields $b(z),\,c(z),\,\bar{b}(\bar{z}),\,\bar{c}(\bar{z})$
transform as 
\begin{eqnarray}
c(z) & \to & e^{2\pi i\alpha_{\mu}^{\prime}}c(z)\,,\nonumber \\
b(z) & \to & e^{2\pi i\alpha_{\mu}^{\prime}}b(z)\,,\nonumber \\
\bar{c}(\bar{z}) & \to & e^{2\pi i\alpha_{\mu}^{\prime}}\bar{c}(\bar{z})\,,\nonumber \\
\bar{b}(\bar{z}) & \to & e^{2\pi i\alpha_{\mu}^{\prime}}\bar{b}(\bar{z})\,,
\end{eqnarray}
if $z$ is moved around the $a_{\mu}\ (\mu=1,\cdots,g)$ cycle once,
and they transform as 
\begin{eqnarray}
c(z) & \to & e^{2\pi i\alpha_{\mu}^{\prime\prime}}c(z)\,,\nonumber \\
b(z) & \to & e^{2\pi i\alpha_{\mu}^{\prime\prime}}b(z)\,,\nonumber \\
\bar{c}(\bar{z}) & \to & e^{2\pi i\alpha_{\mu}^{\prime\prime}}\bar{c}(\bar{z})\,,\nonumber \\
\bar{b}(\bar{z}) & \to & e^{2\pi i\alpha_{\mu}^{\prime\prime}}\bar{b}(\bar{z})\,,
\end{eqnarray}
if $z$ is moved around the $b_{\mu}$ cycle once. We take $\alpha^{\prime}=\alpha^{\prime\prime}=0$
for $\lambda\in\mathbb{Z}$.

Taking the metric on the worldsheet to be the Arakelov metric $g_{z_{i}\bar{z}_{i}}^{A}$,
the correlation function $\left\langle \prod_{i}e^{\epsilon q_{i}\phi}\left(z_{i},\bar{z}_{i}\right)\right\rangle $
is evaluated in \cite{Sonoda1987c} to be 
\begin{equation}
\left\langle \prod_{i}e^{\epsilon q_{i}\phi}\left(z_{i},\bar{z}_{i}\right)\right\rangle \prod_{i}\left(g_{z_{i}\bar{z}_{i}}^{A}\right)^{d\left(q_{i}\right)}\propto\left[\left(\frac{\det^{\prime}\left(-g^{\mathrm{A}z\bar{z}}\partial_{z}\partial_{\bar{z}}\right)}{\det\mathrm{Im}\Omega\int d^{2}z\sqrt{g^{\mathrm{A}}}}\right)^{-\frac{1}{2}}\left|\vartheta{a\atopwithdelims[]b}\left(0|\Omega\right)\right|^{2}\prod_{i>j}e^{-q_{i}q_{j}G^{\mathrm{A}}\left(z_{i},z_{j}\right)}\right]^{\epsilon}\,,\label{eq:leftright}
\end{equation}
where 
\begin{equation}
d\left(q\right)=\frac{1}{2}\epsilon q\left(q+1-2\lambda\right)\,,
\end{equation}
and the characteristics ${a\atopwithdelims[]b}$ is defined so that
\begin{equation}
\left(\Omega a+b\right)_{\nu}=-\sum_{i}q_{i}\int_{P_{0}}^{z_{i}}\omega_{\nu}-\left(2\lambda-1\right)\int_{P_{0}}^{\bigtriangleup}\omega_{\nu}+\left(\Omega\alpha^{\prime}+\alpha^{\prime\prime}\right)_{\nu}\,.
\end{equation}
Here $\Delta$ is the Riemann class, which is related to the canonical
divisor $K$ of the Riemann surface by $2\bigtriangleup=K$.

\subsection{A formula for the superghosts}

Since the correlation function (\ref{eq:leftright}) is left-right
symmetric, we need some more work to get a formula which is useful
for superghosts. We will present it in a form factorized in chiral
and anti-chiral parts. By doing so, it is possible to get the correlation
functions which are not left-right symmetric with respect to the choice
of local operators $e^{q\phi}$ and spin structure.

Substituting (\ref{eq:propagator2}), (\ref{eq:G-F-g-g}) and 
\begin{equation}
\vartheta{a\atopwithdelims[]b}\left(0|\Omega\right)=e^{\pi ia\Omega a+2\pi iab-\pi i\alpha^{\prime}\Omega\alpha^{\prime}-2\pi i\alpha^{\prime}\left(e+\alpha^{\prime\prime}\right)}\vartheta{\alpha^{\prime}\atopwithdelims[]\alpha^{\prime\prime}}\left(e|\Omega\right)\,,
\end{equation}
with 
\begin{equation}
e_{\nu}=-\sum_{i}q_{i}\int_{P_{0}}^{z_{i}}\omega_{\nu}-\left(2\lambda-1\right)\int_{P_{0}}^{\bigtriangleup}\omega_{\nu}\,,
\end{equation}
into the right hand side of (\ref{eq:leftright}), we obtain for $g\ne1$
\begin{eqnarray}
 &  & \left\langle \prod_{i}e^{\epsilon q_{i}\phi}\left(z_{i},\bar{z}_{i}\right)\right\rangle \nonumber \\
 &  & \quad=\left[\left(\frac{\det^{\prime}\left(-g^{\mathrm{A}z\bar{z}}\partial_{z}\partial_{\bar{z}}\right)}{\det\mathrm{Im}\Omega\int d^{2}z\sqrt{g^{\mathrm{A}}}}\right)^{-\frac{1}{2}}\left|\vartheta{\alpha^{\prime}\atopwithdelims[]\alpha^{\prime\prime}}\left(e|\Omega\right)\right|^{2}\prod_{i>j}\left|E\left(z_{i},z_{j}\right)\right|^{2q_{i}q_{j}}\vphantom{\left(\left(g_{z_{i}\bar{z}_{i}}^{A}\right)^{\frac{g}{2}}\exp\left[-\frac{2\pi}{g-1}\mathrm{Im}\int_{\left(g-1\right)z_{i}}^{\bigtriangleup}\omega\left(\mathrm{Im}\Omega\right)^{-1}\mathrm{Im}\int_{\left(g-1\right)z_{i}}^{\bigtriangleup}\omega\right]\right)^{-q_{i}\left(2\lambda-1\right)}}\right.\nonumber \\
 &  & \hphantom{=\quad}\times\left.\prod_{i}\left(\left(g_{z_{i}\bar{z}_{i}}^{A}\right)^{\frac{g}{2}}\exp\left[-\frac{2\pi}{g-1}\mathrm{Im}\int_{\left(g-1\right)z_{i}}^{\bigtriangleup}\omega\left(\mathrm{Im}\Omega\right)^{-1}\mathrm{Im}\int_{\left(g-1\right)z_{i}}^{\bigtriangleup}\omega\right]\right)^{-q_{i}\left(2\lambda-1\right)}\right]^{\epsilon}.~~~~~~~~~\label{eq:leftright2}
\end{eqnarray}
Using (\ref{eq:ArakelovR}), we can see 
\begin{equation}
\partial_{z}\partial_{\bar{z}}\ln\left(\left(g_{z\bar{z}}^{A}\right)^{\frac{g}{2}}\exp\left[-\frac{2\pi}{g-1}\mathrm{Im}\int_{\left(g-1\right)z}^{\bigtriangleup}\omega\left(\mathrm{Im}\Omega\right)^{-1}\mathrm{Im}\int_{\left(g-1\right)z}^{\bigtriangleup}\omega\right]\right)=0\,.
\end{equation}
Therefore there exists a holomorphic $\frac{g}{2}$ form $\sigma\left(z\right)$
such that 
\begin{equation}
\left(g_{z\bar{z}}^{A}\right)^{\frac{g}{2}}\exp\left[-\frac{2\pi}{g-1}\mathrm{Im}\int_{\left(g-1\right)z}^{\bigtriangleup}\omega\left(\mathrm{Im}\Omega\right)^{-1}\mathrm{Im}\int_{\left(g-1\right)z}^{\bigtriangleup}\omega\right]=\left|\sigma\left(z\right)\right|^{2}e^{\frac{3}{g-1}S}\,,\label{eq:sigma}
\end{equation}
where $S$ is independent of $z$. $\sigma\left(z\right)$ has no
zeros or poles, and it should transform as 
\begin{equation}
\sigma\left(z\right)\to e^{-2\pi i\int_{\left(g-1\right)z}^{\bigtriangleup}\omega_{\mu}+\pi i\left(g-1\right)\Omega_{\mu\mu}}\sigma\left(z\right)\,,
\end{equation}
when $z$ is moved around the $b_{\mu}$ cycle once, and invariant
when $z$ is moved around the $a_{\mu}$ cycles. These properties
fix $\sigma\left(z\right)$ and it should coincide with the $\sigma\left(z\right)$
in \cite{Verlinde:1986kw,D'Hoker:1988ta} up to a multiplicative factor.
Substituting (\ref{eq:sigma}) into (\ref{eq:leftright2}), we obtain
\begin{eqnarray}
 &  & \left\langle \prod_{i}e^{\epsilon q_{i}\phi}\left(z_{i},\bar{z}_{i}\right)\right\rangle \nonumber \\
 &  & \quad=\left[\left(\frac{\det^{\prime}\left(-g^{\mathrm{A}z\bar{z}}\partial_{z}\partial_{\bar{z}}\right)}{\det\mathrm{Im}\Omega\int d^{2}z\sqrt{g^{\mathrm{A}}}}\right)^{-\frac{1}{2}}\left|\vartheta{\alpha^{\prime}\atopwithdelims[]\alpha^{\prime\prime}}\left(e|\Omega\right)\prod_{i>j}E\left(z_{i},z_{j}\right)^{q_{i}q_{j}}\prod_{i}\sigma\left(z_{i}\right)^{-q_{i}\left(2\lambda-1\right)}\right|^{2}\right]^{\epsilon}\nonumber \\
 &  & \hphantom{\quad=\quad}\times\ e^{3\left(2\lambda-1\right)^{2}\epsilon S}\,.~~~~~\label{eq:leftright3}
\end{eqnarray}
For $g=1$, instead of (\ref{eq:leftright2}) we get 
\begin{eqnarray}
 &  & \left\langle \prod_{i}e^{\epsilon q_{i}\phi}\left(z_{i},\bar{z}_{i}\right)\right\rangle \nonumber \\
 &  & \quad=\left[\left(\frac{\det^{\prime}\left(-g^{\mathrm{A}z\bar{z}}\partial_{z}\partial_{\bar{z}}\right)}{\det\mathrm{Im}\Omega\int d^{2}z\sqrt{g^{\mathrm{A}}}}\right)^{-\frac{1}{2}}\left|\vartheta{\alpha^{\prime}\atopwithdelims[]\alpha^{\prime\prime}}\left(e|\Omega\right)\right|^{2}\prod_{i>j}\left|E\left(z_{i},z_{j}\right)\right|^{2q_{i}q_{j}}\right]^{\epsilon}\nonumber \\
 &  & \hphantom{\quad=\quad}\times\prod_{i}\left(\left(g_{z_{i}\bar{z}_{i}}^{A}\right)^{\frac{1}{2}}\exp\left[4\pi\mathrm{Im}\int_{P_{0}}^{z_{i}}\omega\left(\mathrm{Im}\Omega\right)^{-1}\mathrm{Im}\int_{P_{0}}^{\bigtriangleup}\omega\right]\right)^{-q_{i}\epsilon\left(2\lambda-1\right)}\nonumber \\
 &  & \hphantom{\quad=\quad}\times\exp\left[-2\pi\left(2\lambda-1\right)^{2}\epsilon\mathrm{Im}\int_{P_{0}}^{\bigtriangleup}\omega\left(\mathrm{Im}\Omega\right)^{-1}\mathrm{Im}\int_{P_{0}}^{\bigtriangleup}\omega\right]\,.
\end{eqnarray}
Since 
\begin{eqnarray}
-2\partial\bar{\partial}\ln g_{z\bar{z}}^{\mathrm{A}} & = & -8\pi(g-1)\mu_{z\bar{z}}=0\,,\nonumber \\
\sum_{i}q_{i} & = & -\left(2\lambda-1\right)\left(g-1\right)=0\,,
\end{eqnarray}
putting 
\begin{eqnarray}
\left(g_{z\bar{z}}^{A}\right)^{\frac{g}{2}}\exp\left[4\pi\mathrm{Im}\int_{P_{0}}^{z}\omega\left(\mathrm{Im}\Omega\right)^{-1}\mathrm{Im}\int_{P_{0}}^{\bigtriangleup}\omega\right] & \equiv & \left|\sigma\left(z\right)\right|^{2}e^{A}\,,\nonumber \\
\exp\left[-2\pi\left(2\lambda-1\right)^{2}\mathrm{Im}\int_{P_{0}}^{\bigtriangleup}\omega\left(\mathrm{Im}\Omega\right)^{-1}\mathrm{Im}\int_{P_{0}}^{\bigtriangleup}\omega\right] & \equiv & e^{3\left(2\lambda-1\right)^{2}S}\,,
\end{eqnarray}
with $S,A$ independent of $z$, we get (\ref{eq:leftright3}).

In (\ref{eq:leftright3}), the correlation function factorizes into
the left and right parts except for the determinant factor and $e^{-3\left(2\lambda-1\right)^{2}S}$.
The determinant factor can also be recast into a factorized form as
follows. Let us consider the $bc$-system with $\lambda=1,\epsilon=1$.
For arbitrary $R,z_{i}\ \left(i=1,\ldots,g\right)$, (\ref{eq:leftright3})
implies 
\begin{eqnarray}
 &  & \left\langle e^{\phi}\left(R,\bar{R}\right)\prod_{i=1}^{g}e^{-\phi}\left(z_{i},\bar{z}_{i}\right)\right\rangle =\frac{\det^{\prime}\left(-g^{\mathrm{A}z\bar{z}}\partial_{z}\partial_{\bar{z}}\right)}{\det\mathrm{Im}\Omega\int d^{2}z\sqrt{g^{\mathrm{A}}}}\left|\det\omega_{\nu z_{i}}\right|^{2}\nonumber \\
 &  & \hspace{2em}=\left(\frac{\det^{\prime}\left(-g^{\mathrm{A}z\bar{z}}\partial_{z}\partial_{\bar{z}}\right)}{\det\mathrm{Im}\Omega\int d^{2}z\sqrt{g^{\mathrm{A}}}}\right)^{-\frac{1}{2}}\left|\vartheta{0\atopwithdelims[]0}\left(e|\Omega\right)\frac{\prod_{i>j}E\left(z_{i},z_{j}\right)}{\prod_{i}E\left(z_{i},R\right)}\frac{\prod_{i}\sigma\left(z_{i}\right)}{\sigma\left(R\right)}\right|^{2}e^{3S}\,,
\end{eqnarray}
and we get 
\begin{equation}
\left(\frac{\det^{\prime}\left(-g^{\mathrm{A}z\bar{z}}\partial_{z}\partial_{\bar{z}}\right)}{\det\mathrm{Im}\Omega\int d^{2}z\sqrt{g^{\mathrm{A}}}}\right)^{-\frac{1}{2}}=\left|\left(\frac{\prod_{i}E\left(z_{i},R\right)\sigma\left(R\right)\det\omega_{\nu z_{i}}}{\vartheta{0\atopwithdelims[]0}\left(e|\Omega\right)\prod_{i>j}E\left(z_{i},z_{j}\right)\prod_{i}\sigma\left(z_{i}\right)}\right)^{\frac{1}{3}}\right|^{2}e^{-S}\,.\label{eq:ZXg}
\end{equation}
Therefore (\ref{eq:leftright3}) can be rewritten as 
\begin{eqnarray}
\left\langle \prod_{i}e^{\epsilon q_{i}\phi}\left(z_{i},\bar{z}_{i}\right)\right\rangle \  & = & \left|\left(\frac{\prod_{i}E\left(z_{i},R\right)\sigma\left(R\right)\det\omega_{\nu z_{i}}}{\vartheta{0\atopwithdelims[]0}\left(e|\Omega\right)\prod_{i>j}E\left(z_{i},z_{j}\right)\prod_{i}\sigma\left(z_{i}\right)}\right)^{\frac{1}{3}}\right|^{2\epsilon}\nonumber \\
 &  & \quad\times\left|\vartheta{\alpha^{\prime}\atopwithdelims[]\alpha^{\prime\prime}}\left(e|\Omega\right)\prod_{i>j}E\left(z_{i},z_{j}\right)^{q_{i}q_{j}}\prod_{i}\sigma\left(z_{i}\right)^{-q_{i}\left(2\lambda-1\right)}\right|^{2\epsilon}\nonumber \\
 &  & \quad\times e^{-\epsilon\left(-3\left(2\lambda-1\right)^{2}+1\right)S}\,.\label{eq:holomorphic}
\end{eqnarray}
$\epsilon\left(-3\left(2\lambda-1\right)^{2}+1\right)$ coincides
with the central charge of the $bc$-system and $e^{-\epsilon\left(-3\left(2\lambda-1\right)^{2}+1\right)S}$
can be identified with the holomorphic anomaly. One can construct
the vacuum amplitude of critical string theory, combining these correlation
functions. With vanishing central charge, the vacuum amplitude completely
factorizes into the holomorphic and antiholomorphic parts.

The holomorphic and antiholomorphic parts of the correlation function
can be read off from (\ref{eq:leftright3}) and we can get the correlation
functions which are not left-right symmetric by combining the holomorphic
and antiholomorphic parts. For example, the partition function of
a free Dirac fermion with spin structure $[\alpha_{L}]$ for left
and $[\alpha_{R}]$ for right can be given by 
\begin{equation}
\left(Z^{\psi}[g_{z\bar{z}}^{\mathrm{A}}]\right)^{2}=\left(\frac{\det^{\prime}\left(-g^{\mathrm{A}z\bar{z}}\partial_{z}\partial_{\bar{z}}\right)}{\det\mathrm{Im}\Omega\int d^{2}z\sqrt{g^{\mathrm{A}}}}\right)^{-\frac{1}{2}}\vartheta[\alpha_{L}]\left(0|\Omega\right)\vartheta[\alpha_{R}]\left(0|\Omega\right)^{*}\,.\label{eq:Zpsi}
\end{equation}
The correlation function 
\[
\int\left[d\beta d\bar{\beta}d\gamma d\bar{\gamma}\right]e^{-S_{\beta\gamma}}\prod_{I}\left|\delta\left(\beta\left(z_{I}\right)\right)\right|^{2}\prod_{r}\left|\delta\left(\gamma\left(Z_{r}\right)\right)\right|^{2}
\]
of the superreparametrization ghost with spin structure $[\alpha_{L}]$
for left and $[\alpha_{R}]$ for right can be evaluated to be 
\begin{eqnarray}
 &  & \left(\frac{\det^{\prime}\left(-g^{\mathrm{A}z\bar{z}}\partial_{z}\partial_{\bar{z}}\right)}{\det\mathrm{Im}\Omega\int d^{2}z\sqrt{g^{\mathrm{A}}}}\right)^{\frac{1}{2}}\left(\vartheta[\alpha_{L}]\left(\left.-\sum_{r}\int_{P_{0}}^{Z_{r}}\omega+\sum_{I}\int_{P_{0}}^{z_{I}}\omega-2\int_{P_{0}}^{\bigtriangleup}\omega\right|\Omega\right)\right)^{-1}\nonumber \\
 &  & \hspace{9em}\times\left(\vartheta[\alpha_{R}]\left(\left.-\sum_{r}\int_{P_{0}}^{Z_{r}}\omega+\sum_{I}\int_{P_{0}}^{z_{I}}\omega-2\int_{P_{0}}^{\bigtriangleup}\omega\right|\Omega\right)^{*}\right)^{-1}\nonumber \\
 &  & \hspace{9em}\times\left|\frac{\prod_{I,r}E\left(z_{I},Z_{r}\right)}{\prod_{I>J}E\left(z_{I},z_{J}\right)\prod_{r>s}E\left(Z_{r},Z_{s}\right)}\frac{\prod_{r}\sigma\left(Z_{r}\right)^{2}}{\prod_{I}\sigma\left(z_{I}\right)^{2}}\right|^{2}e^{-12S}\,.
\end{eqnarray}
Since $Z_{r}\,\left(r=1,\ldots N\right)$ and $z_{I}\,\left(I=1,\ldots,2g-2+N\right)$
are the zeros and the poles of the meromorphic one-form $\partial\rho\left(z\right)dz$
respectively, 
\[
-\sum_{r}Z_{r}+\sum_{I}z_{I}=K=2\bigtriangleup
\]
holds in the divisor sense. Therefore we obtain 
\begin{eqnarray}
 &  & \left(Z^{\psi}[g_{z\bar{z}}^{\mathrm{A}}]\right)^{2}\int\left[d\beta d\gamma\right]e^{-S_{\beta\gamma}}\prod_{I}\left|\delta\left(\beta\left(z_{I}\right)\right)\right|^{2}\prod_{r}\left|\delta\left(\gamma\left(Z_{r}\right)\right)\right|^{2}\nonumber \\
 &  & \qquad\quad=\left|\frac{\prod_{I,r}E\left(z_{I},Z_{r}\right)}{\prod_{I>J}E\left(z_{I},z_{J}\right)\prod_{r>s}E\left(Z_{r},Z_{s}\right)}\frac{\prod_{r}\sigma\left(Z_{r}\right)^{2}}{\prod_{I}\sigma\left(z_{I}\right)^{2}}\right|^{2}e^{-12S}\,.
\end{eqnarray}
On the other hand, (\ref{eq:propagator2}), (\ref{eq:G-F-g-g}) and
(\ref{eq:sigma}) imply 
\begin{eqnarray}
 &  & \left|\frac{\prod_{I,r}E\left(z_{I},Z_{r}\right)}{\prod_{I>J}E\left(z_{I},z_{J}\right)\prod_{r>s}E\left(Z_{r},Z_{s}\right)}\frac{\prod_{r}\sigma\left(Z_{r}\right)^{2}}{\prod_{I}\sigma\left(z_{I}\right)^{2}}\right|^{2}e^{-12S}\nonumber \\
 &  & \qquad=\exp\left[\sum_{I<J}G^{\mathrm{A}}\left(z_{I};z_{J}\right)+\sum_{r<s}G^{\mathrm{A}}\left(Z_{r};Z_{s}\right)-\sum_{I,r}G^{\mathrm{A}}\left(z_{I};Z_{r}\right)\right]\nonumber \\
 &  & \hphantom{\qquad=}\quad\times\prod_{I}\left(2g_{z_{I}\bar{z}_{I}}^{\mathrm{A}}\right)^{-\frac{3}{2}}\prod_{r}\left(2g_{Z_{r}\bar{Z}_{r}}^{\mathrm{A}}\right)^{\frac{1}{2}}\,,
\end{eqnarray}
and from (\ref{eq:e^-Gamma}) we get 
\begin{eqnarray}
 &  & \prod_{r=1}^{N}e^{-\mathop{\mathrm{Re}}\bar{N}_{00}^{rr}}\prod_{I}\left|\partial^{2}\rho\left(z_{I}\right)\right|^{-\frac{3}{2}}e^{\frac{1}{2}\Gamma\left[g_{z\bar{z}}^{\mathrm{A}},\,\rho+\bar{\rho}\right]}\left(Z^{\psi}[g_{z\bar{z}}^{\mathrm{A}}]\right)^{-2}\nonumber \\
 &  & \quad=\int\left[d\beta d\gamma\right]e^{-S_{\beta\gamma}}\prod_{I}\left|\delta\left(\beta\left(z_{I}\right)\right)\right|^{2}\prod_{r}\left|\delta\left(\gamma\left(Z_{r}\right)\right)\right|^{2}\label{eq:superghostZpsi}
\end{eqnarray}
This identity is crucial for deriving the BRST invariant expression
of the superstring amplitudes.

\subsection{A formula for the reparametrization ghosts}

\begin{figure}[h]
\begin{centering}
\includegraphics{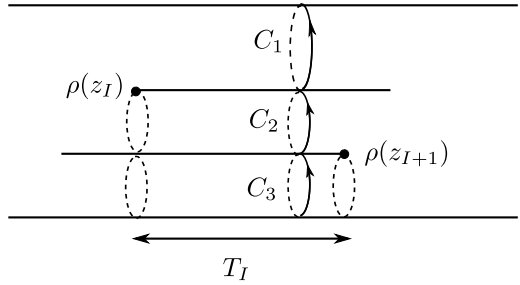} 
\par\end{centering}
\caption{$C_{1},C_{2},C_{3}$ are the contours corresponding to the variation
$T_{I}\to T_{I}+\delta T_{I}$\label{fig:stretch}}
\end{figure}

In \cite{Ishibashi:2013nma} it was shown that the following identity
holds:

\begin{eqnarray}
 &  & \prod_{r=1}^{N}\left(\alpha_{r}e^{2\mathop{\mathrm{Re}}\bar{N}_{00}^{rr}}\right)e^{-\Gamma\left[g_{z\bar{z}}^{\mathrm{A}},\,\rho+\bar{\rho}\right]}\left(Z^{X}[g_{z\bar{z}}^{\mathrm{A}}]\right)^{-2}\nonumber \\
 &  & =\mbox{const.}\int\left[dbd\bar{b}dcd\bar{c}\right]_{g_{z\bar{z}}^{\mathrm{A}}}e^{-S^{bc}}\prod_{r=1}^{N}c\bar{c}(Z_{r},\bar{Z}_{r})\prod_{K=1}^{6g-6+2N}\left[\int dz\wedge d\bar{z}\,i\left(\mu_{K}b+\bar{\mu}_{K}\bar{b}\right)\right].~~~~~~~\label{eq:ghostZX}
\end{eqnarray}
Here $S^{bc}$ is the action for the reparametrization ghosts, $\mu_{K}\,\left(K=1,\ldots,6g-6+2N\right)$
denote the Beltrami differentials for the moduli parameters $T,\alpha,\theta$,
and $\mbox{const.}$ indicates a constant independent of the moduli
parameters. The antighost insertion $\int dz\wedge d\bar{z}\,i\left(\mu_{K}b+\bar{\mu}_{K}\bar{b}\right)$
corresponding to the variations of $T,\alpha,\theta$ are given by
the following contour integrals: 
\begin{itemize}
\item The stretch corresponds to the variation $T\to T+\delta T$ of the
height $T$ of cylinders. Let us order the interaction points $z_{I}\,(I=1,\ldots,2g-2+N)$
so that 
\begin{equation}
\mathrm{Re}\rho(z_{1})\leq\mathrm{Re}\rho(z_{2})\leq\cdots\leq\mathrm{Re}\rho(z_{2g-2+N})\,,
\end{equation}
and define the moduli parameters corresponding to the heights as 
\begin{equation}
T_{I}\equiv\mathrm{Re}\rho(z_{I+1})-\mathrm{Re}\rho(z_{I})\qquad(I=1,\ldots,2g-3+N)\,.
\end{equation}
The antighost insertion corresponding to the deformation $T_{I}\to T_{I}+\delta T_{I}$
is given by 
\begin{equation}
\sum_{i}\oint_{C_{i}}d\left(\mathrm{Im}\rho\right)\left(b_{\rho\rho}+b_{\bar{\rho}\bar{\rho}}\right)=-i\sum_{i}\left(\oint_{C_{i}}\frac{dz}{\partial\rho}b_{zz}-\oint_{\bar{C}_{i}}\frac{d\bar{z}}{\bar{\partial}\bar{\rho}}b_{\bar{z}\bar{z}}\right)\,,\label{eq:stretch}
\end{equation}
where $C_{i}$ denotes the contour around a cylinder which includes
the region $\mathrm{Re}\rho(z_{I})\leq\mathrm{Re}\rho\leq\mathrm{Re}\rho(z_{I+1})$
and the sum should be taken over all such contours. (See figure \ref{fig:stretch}.)
There are $2g-3+N$ insertions of this kind. 
\end{itemize}

\begin{figure}[h]
\begin{centering}
\includegraphics{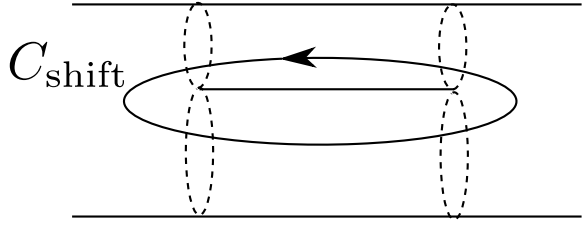} 
\par\end{centering}
\caption{$C_{\mathrm{shift}}$ corresponds to the variation of the momenta
$p^{+}$ flowing along it.\label{fig:shift} }
\end{figure}

\begin{itemize}
\item The twist $\theta\to\theta+\delta\theta$ corresponds to the rotation
of one end of a cylinder with respect to the other. The antighost
insertion should be 
\begin{equation}
\oint_{C_{\mathrm{twist}}}d\left(\mathrm{Im}\rho\right)\left(b_{\rho\rho}-b_{\bar{\rho}\bar{\rho}}\right)=-i\left(\oint_{C_{\mathrm{twist}}}\frac{dz}{\partial\rho}b_{zz}+\oint_{\bar{C}_{\mathrm{twist}}}\frac{d\bar{z}}{\bar{\partial}\bar{\rho}}b_{\bar{z}\bar{z}}\right)\,,\label{eq:twist}
\end{equation}
where $C_{\mathrm{twist}}$ is the contour around the cylinder which
is twisted. There are $3g-3+N$ insertions of this kind. 
\item The shift corresponds to the variation of the loop momenta preserving
the momenta of the external lines. The antighost insertion for such
a variation becomes 
\begin{equation}
-i\left(\oint_{C_{\mathrm{shift}}}\frac{dz}{\partial\rho}b_{zz}+\oint_{\bar{C}_{\mathrm{shift}}}\frac{d\bar{z}}{\bar{\partial}\bar{\rho}}b_{\bar{z}\bar{z}}\right)\,,\label{eq:shift}
\end{equation}
where $C_{\mathrm{shift}}$ is taken as in figure \ref{fig:shift}.
There are $g$ insertions of this kind. 
\end{itemize}

Therefore (\ref{eq:ghostZX}) can be expressed in terms of the contour
integrals as 
\begin{eqnarray}
 &  & \prod_{r=1}^{N}\left(\alpha_{r}e^{2\mathop{\mathrm{Re}}\bar{N}_{00}^{rr}}\right)e^{-\Gamma\left[g_{z\bar{z}}^{\mathrm{A}},\,\rho+\bar{\rho}\right]}\left(Z^{X}[g_{z\bar{z}}^{\mathrm{A}}]\right)^{-2}\nonumber \\
 &  & =\mbox{const.}\int\left[dbd\bar{b}dcd\bar{c}\right]_{g_{z\bar{z}}^{\mathrm{A}}}e^{-S^{bc}}\prod_{r=1}^{N}c\bar{c}(Z_{r},\bar{Z}_{r})\prod_{K=1}^{6g-6+2N}\left[\oint_{C_{K}}\frac{dz}{\partial\rho}b_{zz}+\varepsilon_{K}\oint_{\bar{C}_{K}}\frac{d\bar{z}}{\bar{\partial}\bar{\rho}}b_{\bar{z}\bar{z}}\right].~~~~~~~\label{eq:ZXcontour}
\end{eqnarray}
Here $\varepsilon_{K}=-1$ for the stretches and $\varepsilon_{K}=1$
for the twists and shifts.

\section{A proof of (\ref{eq:superBRSTinvariant})\label{sec:A-proof-of}}

We would like to prove (\ref{eq:superBRSTinvariant}) by showing that
the right hand side of (\ref{eq:superBRSTinvariant}) is equal to
that of (\ref{eq:superFN3}). We consider the generic situation in
which $z_{I},Z_{r}$ are all distinct.

What we do first is to rewrite the PCO's $X\left(z_{I}\right),\bar{X}\left(\bar{z}_{I}\right)$
using the existence of nilpotent fermionic charge 
\begin{eqnarray}
\hat{Q} & \equiv & \oint\frac{dz}{2\pi i}\partial\rho\left(z\right)\left[c\left(i\partial X^{+}-\frac{1}{2}\partial\rho\right)\left(z\right)+\frac{1}{2}\gamma\psi^{+}\left(z\right)\right]\nonumber \\
 &  & {}+\oint\frac{d\bar{z}}{2\pi i}\bar{\partial}\bar{\rho}\left(\bar{z}\right)\left[\bar{c}\left(i\bar{\partial}X^{+}-\frac{1}{2}\bar{\partial}\bar{\rho}\right)\left(\bar{z}\right)+\frac{1}{2}\bar{\gamma}\bar{\psi}^{+}\left(\bar{z}\right)\right]\,.
\end{eqnarray}
One can show that $X\left(z_{I}\right)$ can be expressed as 
\begin{eqnarray}
X\left(z_{I}\right) & = & -e^{\phi}T_{F}^{\mathrm{LC}}\left(z_{I}\right)+\left\{ \hat{Q},\oint_{z_{I}}\frac{dw}{2\pi i}\frac{1}{w-z_{I}}\mathcal{O}\left(w\right)e^{\phi}\left(z_{I}\right)\right\} \nonumber \\
 &  & \quad+\frac{1}{4}\oint_{z_{I}}\frac{dw}{2\pi i}\frac{1}{w-z_{I}}\partial\rho\psi^{-}\left(w\right)e^{\phi}\left(z_{I}\right)+\frac{1}{4}b\left(\partial\eta e^{2\phi}+\frac{1}{2}\eta\partial e^{2\phi}\right)\left(z_{I}\right)\,,\label{eq:PCOQ}
\end{eqnarray}
where 
\begin{eqnarray}
\mathcal{O} & = & \frac{i}{\partial\rho}\partial X^{-}\beta+\frac{1}{2}\frac{b}{\partial\rho}\psi^{-}\nonumber \\
 &  & \quad+2Q^{2}i\left[\left(\frac{5}{4}\frac{\left(\partial^{2}X^{+}\right)^{2}}{\left(\partial X^{+}\right)^{3}}-\frac{1}{2}\frac{\partial^{3}X^{+}}{\left(\partial X^{+}\right)^{2}}\right)\frac{2\beta}{\partial\rho}\right.\nonumber \\
 &  & \hphantom{\quad+2Q^{2}i\quad}\left.-\frac{2\partial^{2}X^{+}}{\left(\partial X^{+}\right)^{2}}\partial\left(\frac{2\beta}{\partial\rho}\right)+\frac{\partial^{2}\left(\frac{2\beta}{\partial\rho}\right)}{\partial X^{+}}-\frac{\frac{2\beta}{\partial\rho}\partial\psi^{+}\partial^{2}\psi^{+}}{2\left(\partial X^{+}\right)^{3}}\right]\,,
\end{eqnarray}
and $\hat{Q}$ satisfies the following identities: 
\begin{eqnarray}
 &  & \left[\hat{Q},c\bar{c}e^{-\phi-\bar{\phi}}V_{r}^{\mathrm{DDF}}(Z_{r},\bar{Z}_{r})\right]=0\,,\nonumber \\
 &  & \left[\hat{Q},e^{\phi}T_{F}^{\mathrm{LC}}\left(z_{I}\right)\right]=\left[Q,e^{\bar{\phi}}\bar{T}_{F}^{\mathrm{LC}}\left(\bar{z}_{I}\right)\right]=0\,,\nonumber \\
 &  & \left[\hat{Q},\oint_{z_{I}}\frac{dw}{2\pi i}\frac{1}{w-z_{I}}\partial\rho\psi^{-}\left(w\right)e^{\phi}\left(z_{I}\right)\right]=\left[\hat{Q},\oint_{\bar{z}_{I}}\frac{d\bar{w}}{2\pi i}\frac{1}{\bar{w}-\bar{z}_{I}}\bar{\partial}\bar{\rho}\bar{\psi}^{-}\left(\bar{w}\right)e^{\bar{\phi}}\left(\bar{z}_{I}\right)\right]=0\,,\nonumber \\
 &  & \left[\hat{Q},\frac{1}{4}b\left(\partial\eta e^{2\phi}+\frac{1}{2}\eta\partial e^{2\phi}\right)\left(z_{I}\right)\right]=\left[\hat{Q},\frac{1}{4}\bar{b}\left(\bar{\partial}\bar{\eta}e^{2\bar{\phi}}+\frac{1}{2}\bar{\eta}\bar{\partial}e^{2\bar{\phi}}\right)\left(\bar{z}_{I}\right)\right]=0\,,\nonumber \\
 &  & \left\{ \hat{Q},\mathcal{S}\left(\mathbf{z},Z_{r}\right)\right\} =\left\{ \hat{Q},\bar{\mathcal{S}}\left(\bar{\mathbf{z}},\bar{Z}_{r}\right)\right\} =0\,,\nonumber \\
 &  & \left\{ \hat{Q},e^{\phi}\partial\left(\psi^{+}e^{-\frac{Q^{2}}{2}\frac{i}{p_{r}^{+}}X_{L}^{+}}\right)\left(z_{I}\right)\right\} =\left\{ \hat{Q},e^{\bar{\phi}}\partial\left(\bar{\psi}^{+}e^{-\frac{Q^{2}}{2}\frac{i}{p_{r^{\prime}}^{+}}X_{R}^{+}}\right)\left(\bar{z}_{I}\right)\right\} =0\,.
\end{eqnarray}
The antighost insertions ${\displaystyle \prod_{K=1}^{6g-6+2N}\left[\oint_{C_{K}}\frac{dz}{\partial\rho}b_{zz}+\varepsilon_{K}\oint_{\bar{C}_{K}}\frac{d\bar{z}}{\bar{\partial}\bar{\rho}}b_{\bar{z}\bar{z}}\right]}$
is a product of the contour integrals of the types (\ref{eq:stretch}),
(\ref{eq:twist}) and (\ref{eq:shift}). The anticommutator of $\hat{Q}$
with the contour integral of the type (\ref{eq:stretch}) becomes
\begin{eqnarray}
 &  & \left\{ \hat{Q},-i\sum_{i}\left(\oint_{C_{i}}\frac{dz}{\partial\rho}b_{zz}-\oint_{\bar{C}_{i}}\frac{d\bar{z}}{\bar{\partial}\bar{\rho}}b_{\bar{z}\bar{z}}\right)\right\} \nonumber \\
 &  & \quad=-i\sum_{i}\left(\oint_{C_{i}}dz\left(i\partial X^{+}-\frac{1}{2}\partial\rho\right)-\oint_{\bar{C}_{i}}d\bar{z}\left(i\bar{\partial}X^{+}-\frac{1}{2}\bar{\partial}\rho\right)\right)\,.\label{eq:Qstretch}
\end{eqnarray}
Since 
\[
-i\sum_{i}\left(\oint_{C_{i}}dzi\partial X^{+}-\oint_{\bar{C}_{i}}d\bar{z}i\bar{\partial}X^{+}\right)
\]
and 
\[
-i\sum_{i}\left(\oint_{C_{i}}dz\frac{1}{2}\partial\rho-\oint_{\bar{C}_{i}}d\bar{z}\frac{1}{2}\bar{\partial}\rho\right)
\]
are both equal to the total momentum in the $+$ direction through
the channel which is fixed by the external momenta, the right hand
side of (\ref{eq:Qstretch}) should vanish. In the case of the contour
integrals (\ref{eq:twist}), (\ref{eq:shift}), we obtain 
\begin{eqnarray}
 &  & \left\{ \hat{Q},-i\left(\oint_{C}\frac{dz}{\partial\rho}b_{zz}+\oint_{\bar{C}}\frac{d\bar{z}}{\bar{\partial}\bar{\rho}}b_{\bar{z}\bar{z}}\right)\right\} \nonumber \\
 &  & \quad=-i\left(\oint_{C}dz\left(i\partial X^{+}-\frac{1}{2}\partial\rho\right)+\oint_{\bar{C}}d\bar{z}\left(i\bar{\partial}X^{+}-\frac{1}{2}\bar{\partial}\rho\right)\right)\,,
\end{eqnarray}
which vanishes because $X^{+}$and $\rho+\bar{\rho}$ should be singlevalued.
Hence $\hat{Q}$ commutes or anticommutes with all the quantities
in (\ref{eq:superBRSTinvariant2}). Therefore the second term on the
right hand side of (\ref{eq:PCOQ}) does not contribute to the correlation
functions, because it is $\hat{Q}$ exact. We can replace all the
$X\left(z_{I}\right)$ in the correlation functions by 
\begin{equation}
-e^{\phi}T_{F}^{\mathrm{LC}}\left(z_{I}\right)+\frac{1}{4}\oint_{z_{I}}\frac{dw}{2\pi i}\frac{1}{w-z_{I}}\partial\rho\psi^{-}\left(w\right)e^{\phi}\left(z_{I}\right)+\frac{1}{4}b\left(\partial\eta e^{2\phi}+\frac{1}{2}\eta\partial e^{2\phi}\right)\left(z_{I}\right)\,,\label{eq:PCO2}
\end{equation}
and similarly for $\bar{X}\left(\bar{z}_{I}\right)$. Then the third
term in (\ref{eq:PCO2}) can be omitted because of the ghost number
conservation and similarly for the antiholomorphic sector.

Replacing $X\left(z_{I}\right)$ by 
\begin{equation}
-e^{\phi}T_{F}^{\mathrm{LC}}\left(z_{I}\right)+\frac{1}{4}\oint_{z_{I}}\frac{dw}{2\pi i}\frac{1}{w-z_{I}}\partial\rho\psi^{-}\left(w\right)e^{\phi}\left(z_{I}\right)\,,
\end{equation}
and similarly for $\bar{X}\left(\bar{z}_{I}\right)$, the right hand
side of (\ref{eq:superBRSTinvariant}) becomes a sum of the right
hand side of (\ref{eq:superFN3}) and the terms which involve 
\begin{equation}
\oint_{z_{I}}\frac{dw}{2\pi i}\frac{1}{w-z_{I}}\partial\rho\psi^{-}\left(w\right)e^{\phi}\left(z_{I}\right)\,,\qquad\oint_{\bar{z}_{I}}\frac{d\bar{w}}{2\pi i}\frac{1}{\bar{w}-\bar{z}_{I}}\bar{\partial}\bar{\rho}\bar{\psi}^{-}\left(\bar{w}\right)e^{\bar{\phi}}\left(\bar{z}_{I}\right)\,,\label{eq:psi-}
\end{equation}
or 
\begin{equation}
e^{\phi}\partial\left(\psi^{+}e^{-\frac{Q^{2}}{2}\frac{i}{p_{r}^{+}}X_{L}^{+}}\right)\left(z_{I}\right)\,,\qquad e^{\bar{\phi}}\partial\left(\bar{\psi}^{+}e^{-\frac{Q^{2}}{2}\frac{i}{p_{r^{\prime}}^{+}}X_{R}^{+}}\right)\left(\bar{z}_{I}\right)\,,\label{eq:psi+}
\end{equation}
which appear in (\ref{eq:XS1}) and (\ref{eq:XS2}), in place of $e^{\phi}T_{F}^{\mathrm{LC}}\left(z_{I}\right),e^{\bar{\phi}}\bar{T}_{F}^{\mathrm{LC}}\left(\bar{z}_{I}\right)$.
The $X^{\pm}$ CFT part of such terms are of the form 
\begin{eqnarray}
 &  & \Bigl\langle\mathcal{O}^{-}\left(z_{I_{1}}\right)\cdots\mathcal{O}^{-}\left(z_{I_{n}}\right)\mathcal{O}^{+}\left(z_{I_{n+1}}\right)\cdots\mathcal{O}^{+}\left(z_{I_{n+m}}\right)\bar{\mathcal{O}}^{-}\cdots\bar{\mathcal{O}}^{+}\cdots\nonumber \\
 &  & \qquad\quad\times\left(\mbox{contributions from }\mathcal{S},\bar{\mathcal{S}},V_{r}^{\mathrm{DDF}}\right)\Bigr\rangle^{X^{\pm}}\,,\label{eq:uvXpm}
\end{eqnarray}
where 
\begin{eqnarray}
\mathcal{O}^{-}\left(z_{I}\right) & \equiv & \oint_{z_{I}}\frac{dw}{2\pi i}\frac{1}{w-z_{I}}\partial\rho\psi^{-}\left(w\right)\,,\nonumber \\
\mathcal{O}^{+}\left(z_{I}\right) & \equiv & \partial\left(\psi^{+}e^{-\frac{Q^{2}}{2}\frac{i}{p_{r}^{+}}X_{L}^{+}}\right)\left(z_{I}\right)\,,
\end{eqnarray}
and the antiholomorphic versions are defined in a similar way. Here
$z_{I_{1}},\ldots,z_{I_{n+m}}$ are all distinct. We would like to
show that the correlation functions of the form (\ref{eq:uvXpm})
vanish. (\ref{eq:fermioniccorr2}) implies that in calculating correlation
functions of the form (\ref{eq:uvXpm}), all the $\psi^{+}$'s in
$\mathcal{O}^{+}$'s should be contracted with $\psi^{-}$'s, which
come from $\mathcal{O}^{-}\left(z_{I}\right)$. Therefore (\ref{eq:uvXpm})
with $m\ne0$ should involve a factor of the form 
\begin{equation}
\oint_{z_{I}}\frac{dw}{2\pi i}\frac{1}{w-z_{I}}\partial\rho\begC1{\psi^{-}}\conC{\left(w\right)\partial^{k}}\endC1{\psi^{+}}\left(z_{J}\right)\,,\label{eq:uvcontract}
\end{equation}
for some integer $k\geq0$, with $z_{I}\ne z_{J}$. (\ref{eq:uvcontract})
vanishes because the contraction does not have any poles at $w=z_{I}$
and $\partial\rho\left(z_{I}\right)=0$. Therefore the correlation
function (\ref{eq:uvXpm}) vanishes if it involves $\mathcal{O}^{+}$
or $\bar{\mathcal{O}}^{+}$. What remains to be done is to prove the
correlation functions of the form 
\begin{equation}
\left\langle \mathcal{O}^{-}\left(z_{I_{1}}\right)\cdots\mathcal{O}^{-}\left(z_{I_{n}}\right)\bar{\mathcal{O}}^{-}\cdots\times\left(\mbox{contributions from }\mathcal{S},\bar{\mathcal{S}},V_{r}^{\mathrm{DDF}}\right)\right\rangle ^{X^{\pm}}\label{eq:uvXpm2}
\end{equation}
vanish using (\ref{eq:fermioniccorr2}). The contractions of $\psi^{-}$'s
in $\mathcal{O}^{-}$ with $\psi^{+}$'s from $V_{r}^{\mathrm{DDF}}$
do not contribute because 
\[
\oint_{z_{I}}\frac{dw}{2\pi i}\frac{1}{w-z_{I}}\partial\rho\begC1{\psi^{-}}\conC{\left(w\right)\partial^{k}}\endC1{\psi^{+}}\left(Z_{r}\right)=0\,.
\]
The contractions of $\psi^{-}$'s in $\mathcal{O}^{-}$ and $\psi^{+}$'s
from $\mathcal{S}$ inevitably induce a factor of the form 
\[
\oint_{z_{I}}\frac{dw}{2\pi i}\frac{1}{w-z_{I}}\partial\rho\begC1{\psi^{-}}\conC{\left(w\right)\partial^{k}}\endC1{\psi^{+}}\left(z\right)=0\,,
\]
with $z\sim z_{J}\ne z_{I}$, because $\mathcal{S}$'s involves even
number of $\psi^{+}$'s. The contractions of $\psi^{-}$'s in $\mathcal{O}^{-}$
and $\psi^{+}$'s from $S_{\mathrm{int}}$ necessarily induce a factor
of the form 
\[
\oint_{z_{I}}\frac{dw}{2\pi i}\frac{1}{w-z_{I}}\partial\rho\begC1{\psi^{-}}\conC{\left(w\right)\partial^{k}}\endC1{\psi^{+}}\left(z_{J}\right)=0\,,
\]
with $z_{J}\ne z_{I}$, because $S_{\mathrm{int}}$ is localized at
the interaction points and involves even number of $\psi^{+}$'s.

Thus we have shown that the terms of the form (\ref{eq:uvXpm}) all
vanish and the right hand side of (\ref{eq:superBRSTinvariant}) is
equal to that of (\ref{eq:superFN3}). 

\providecommand{\href}[2]{#2}\begingroup\raggedright\endgroup


\end{document}